\documentclass[11pt]{article}
\usepackage{a4wide,psfrag,epsfig,amsfonts,amsmath,amssymb,color,enumerate}

\newtheorem{lemma}{Lemma}

\newtheorem{definition}{Definition}
\newtheorem{theorem}[lemma]{Theorem}

\newtheorem{proposition}[lemma]{Proposition}
\makeatletter

\DeclareRobustCommand{\qed}{%
  \ifmmode
  \else \leavevmode\unskip\penalty9999 \hbox{}\nobreak\hfill
  \fi
  \quad\hbox{\qedsymbol}}
\newcommand{\openbox}{\leavevmode
  \hbox to.77778em{%
  \hfil\vrule
  \vbox to.675em{\hrule width.6em\vfil\hrule}%
  \vrule\hfil}}
\newcommand{\qedsymbol}{\openbox}
\newenvironment{proof}[1][\proofname]{\par
  \normalfont
  \topsep6\p@\@plus6\p@ \trivlist
  \item[\hskip\labelsep\itshape
    #1.]\ignorespaces
}{%
  \qed\endtrivlist
}
\newcommand{\proofname}{Proof}
\newenvironment{remark}[1][{\bf Remark}]{\par
  \normalfont
  \topsep6\p@\@plus6\p@ \trivlist
  \item[\hskip\labelsep\itshape
    #1.]\ignorespaces
}{
\endtrivlist
}

\makeatother

\newcommand{\dtv}{\mathrm{d}_\mathrm{TV}}
\def\zset{\ensuremath{\mathbb{Z}}}

\def\epsilon{\varepsilon}
\def\calM{\mathcal{M}}
\def\calB{\mathcal{B}}
\def\expected{\mathbb{E}}
\def\mon{\text{mon}}
\def\calM{\mathcal{M}}

\newcommand{\ConvexBig}{
\begin{picture}(3,4)

\put(1,1){\line(1,0){1}} \put(2,1){\line(0,1){1}} \put(1,2){\line(1,0){1}} \put(1,1){\line(0,1){1}}

\put(2,2){\line(0,2){1}} \put(1,3){\line(1,0){1}} \put(1,2){\line(0,1){1}}

\put(1,0){\makebox(1,1){\figuresize{$1$/$2$}}}
\put(0,1){\makebox(1,1){\figuresize{$1$}}}
\put(2,1){\makebox(1,1){\figuresize{$2$}}}
\put(0,2){\makebox(1,1){\figuresize{$1$}}}
\put(2,2){\makebox(1,1){\figuresize{$2$}}}
\put(1,3){\makebox(1,1){\figuresize{$1$}}}

\put(1,1){\makebox(1,1){\figuresize{$f_X$}}}
\put(1,2){\makebox(1,1){\figuresize{$y$}}}

\put(1.5,0.8){\line(0,1){.3}}
\put(1.5,2.9){\line(0,1){.3}}
\put(0.8,1.5){\line(1,0){.3}}
\put(0.8,2.5){\line(1,0){.3}}
\put(1.9,1.5){\line(1,0){.3}}
\put(1.9,2.5){\line(1,0){.3}}

\end{picture}
}

\newcommand{\ConvexSmallA}{
\begin{picture}(3,4)

\put(1,1){\line(1,0){1}} \put(2,1){\line(0,1){1}} \put(1,2){\line(1,0){1}} \put(1,1){\line(0,1){1}}

\put(1,0){\makebox(1,1){\figuresize{$1$/$2$}}}
\put(0,1){\makebox(1,1){\figuresize{$1$}}}
\put(2,1){\makebox(1,1){\figuresize{$2$}}}
\put(1,2){\makebox(1,1){\figuresize{$1$}}}

\put(1,1){\makebox(1,1){\figuresize{$f_X$}}}

\put(1.5,0.8){\line(0,1){.3}}
\put(1.5,1.9){\line(0,1){.3}}
\put(0.8,1.5){\line(1,0){.3}}
\put(1.9,1.5){\line(1,0){.3}}

\end{picture}
}

\newcommand{\ConvexSmallB}{
\begin{picture}(3,4)

\put(1,1){\line(1,0){1}} \put(2,1){\line(0,1){1}} \put(1,2){\line(1,0){1}} \put(1,1){\line(0,1){1}}

\put(1,0){\makebox(1,1){\figuresize{$1$/$2$}}}
\put(0,1){\makebox(1,1){\figuresize{$1$}}}
\put(2,1){\makebox(1,1){\figuresize{$2$}}}
\put(1,2){\makebox(1,1){\figuresize{$2$}}}

\put(1,1){\makebox(1,1){\figuresize{$f_X$}}}

\put(1.5,0.8){\line(0,1){.3}}
\put(1.5,1.9){\line(0,1){.3}}
\put(0.8,1.5){\line(1,0){.3}}
\put(1.9,1.5){\line(1,0){.3}}

\end{picture}
}

\newcommand{\Qone}{
\begin{picture}(4,2)
\put(1,1){\line(1,0){1}} \put(1,1){\line(0,1){1}} \put(1,2){\line(1,0){1}} \put(2,1){\line(0,1){1}}
\put(2,1){\line(1,0){1}} \put(2,1){\line(0,1){1}} \put(2,2){\line(1,0){1}} \put(3,1){\line(0,1){1}}
\put(3,1){\line(1,0){1}} \put(3,1){\line(0,1){1}} \put(3,2){\line(1,0){1}} \put(4,1){\line(0,1){1}}
\put(0,0){\line(1,0){1}} \put(0,0){\line(0,1){1}} \put(0,1){\line(1,0){1}} \put(1,0){\line(0,1){1}}
\put(1,0){\line(1,0){1}} \put(1,0){\line(0,1){1}} \put(1,1){\line(1,0){1}} \put(2,0){\line(0,1){1}}
\put(2,0){\line(1,0){1}} \put(2,0){\line(0,1){1}} \put(2,1){\line(1,0){1}} \put(3,0){\line(0,1){1}}\put(2,0){\makebox(1,1){\figuresize{$f_X$}}}
\put(2.50,-0.20){\line(0,1){.3}}
\put(3,0){\line(1,0){1}} \put(3,0){\line(0,1){1}} \put(3,1){\line(1,0){1}} \put(4,0){\line(0,1){1}}
\end{picture}
}

\newcommand{\Qtwo}{
\begin{picture}(3,2)
\put(0,1){\line(1,0){1}} \put(0,1){\line(0,1){1}} \put(0,2){\line(1,0){1}} \put(1,1){\line(0,1){1}}
\put(1,1){\line(1,0){1}} \put(1,1){\line(0,1){1}} \put(1,2){\line(1,0){1}} \put(2,1){\line(0,1){1}}
\put(2,1){\line(1,0){1}} \put(2,1){\line(0,1){1}} \put(2,2){\line(1,0){1}} \put(3,1){\line(0,1){1}}
\put(0,0){\line(1,0){1}} \put(0,0){\line(0,1){1}} \put(0,1){\line(1,0){1}} \put(1,0){\line(0,1){1}}
\put(1,0){\line(1,0){1}} \put(1,0){\line(0,1){1}} \put(1,1){\line(1,0){1}} \put(2,0){\line(0,1){1}}\put(1,0){\makebox(1,1){\figuresize{$f_X$}}}
\put(1.50,-0.20){\line(0,1){.3}}
\put(2,0){\line(1,0){1}} \put(2,0){\line(0,1){1}} \put(2,1){\line(1,0){1}} \put(3,0){\line(0,1){1}}
\end{picture}
}

\newcommand{\Qthree}{
\begin{picture}(3,2)
\put(0,1){\line(1,0){1}} \put(0,1){\line(0,1){1}} \put(0,2){\line(1,0){1}} \put(1,1){\line(0,1){1}}
\put(1,1){\line(1,0){1}} \put(1,1){\line(0,1){1}} \put(1,2){\line(1,0){1}} \put(2,1){\line(0,1){1}}
\put(2,1){\line(1,0){1}} \put(2,1){\line(0,1){1}} \put(2,2){\line(1,0){1}} \put(3,1){\line(0,1){1}}
\put(0,0){\line(1,0){1}} \put(0,0){\line(0,1){1}} \put(0,1){\line(1,0){1}} \put(1,0){\line(0,1){1}}
\put(1,0){\line(1,0){1}} \put(1,0){\line(0,1){1}} \put(1,1){\line(1,0){1}} \put(2,0){\line(0,1){1}}\put(1,0){\makebox(1,1){\figuresize{$f_X$}}}
\put(1.50,-0.20){\line(0,1){.3}}
\end{picture}
}

\newcommand{\Qfour}{
\begin{picture}(3,2)
\put(0,1){\line(1,0){1}} \put(0,1){\line(0,1){1}} \put(0,2){\line(1,0){1}} \put(1,1){\line(0,1){1}}
\put(1,1){\line(1,0){1}} \put(1,1){\line(0,1){1}} \put(1,2){\line(1,0){1}} \put(2,1){\line(0,1){1}}
\put(0,0){\line(1,0){1}} \put(0,0){\line(0,1){1}} \put(0,1){\line(1,0){1}} \put(1,0){\line(0,1){1}}
\put(1,0){\line(1,0){1}} \put(1,0){\line(0,1){1}} \put(1,1){\line(1,0){1}} \put(2,0){\line(0,1){1}}\put(1,0){\makebox(1,1){\figuresize{$f_X$}}}
\put(1.50,-0.20){\line(0,1){.3}}
\put(2,0){\line(1,0){1}} \put(2,0){\line(0,1){1}} \put(2,1){\line(1,0){1}} \put(3,0){\line(0,1){1}}
\end{picture}
}

\newcommand{\Qfive}{
\begin{picture}(2,2)
\put(0,1){\line(1,0){1}} \put(0,1){\line(0,1){1}} \put(0,2){\line(1,0){1}} \put(1,1){\line(0,1){1}}
\put(1,1){\line(1,0){1}} \put(1,1){\line(0,1){1}} \put(1,2){\line(1,0){1}} \put(2,1){\line(0,1){1}}
\put(0,0){\line(1,0){1}} \put(0,0){\line(0,1){1}} \put(0,1){\line(1,0){1}} \put(1,0){\line(0,1){1}}
\put(1,0){\line(1,0){1}} \put(1,0){\line(0,1){1}} \put(1,1){\line(1,0){1}} \put(2,0){\line(0,1){1}}\put(1,0){\makebox(1,1){\figuresize{$f_X$}}}
\put(1.50,-0.20){\line(0,1){.3}}
\end{picture}
}

\newcommand{\Qsix}{
\begin{picture}(3,2)
\put(1,1){\line(1,0){1}} \put(1,1){\line(0,1){1}} \put(1,2){\line(1,0){1}} \put(2,1){\line(0,1){1}}
\put(0,0){\line(1,0){1}} \put(0,0){\line(0,1){1}} \put(0,1){\line(1,0){1}} \put(1,0){\line(0,1){1}}
\put(1,0){\line(1,0){1}} \put(1,0){\line(0,1){1}} \put(1,1){\line(1,0){1}} \put(2,0){\line(0,1){1}}\put(1,0){\makebox(1,1){\figuresize{$f_X$}}}
\put(1.50,-0.20){\line(0,1){.3}}
\put(2,0){\line(1,0){1}} \put(2,0){\line(0,1){1}} \put(2,1){\line(1,0){1}} \put(3,0){\line(0,1){1}}
\end{picture}
}

\newcommand{\Qseven}{
\begin{picture}(2,1)
\put(0,0){\line(1,0){1}} \put(0,0){\line(0,1){1}} \put(0,1){\line(1,0){1}} \put(1,0){\line(0,1){1}}
\put(1,0){\line(1,0){1}} \put(1,0){\line(0,1){1}} \put(1,1){\line(1,0){1}} \put(2,0){\line(0,1){1}}\put(1,0){\makebox(1,1){\figuresize{$f_X$}}}
\put(1.50,-0.20){\line(0,1){.3}}
\end{picture}
}

\newcommand{\Vfig}{
\begin{picture}(3,3)
\put(1,1){\line(1,0){1}} \put(1,1){\line(0,1){1}} \put(1,2){\line(1,0){1}} \put(2,1){\line(0,1){1}}\put(1,1){\makebox(1,1){\figuresize{$f_X$}}}
\put(1.50,0.80){\line(0,1){.3}}
\put(0,0){\line(1,0){1}} \put(0,0){\line(0,1){1}} \put(0,1){\line(1,0){1}} \put(1,0){\line(0,1){1}} \put(0,0){\line(1,1){1}} \put(1,0){\line(-1,1){1}}
\end{picture}
}

\newcommand{\Wfig}{
\begin{picture}(3,3)
\put(1,1){\line(1,0){1}} \put(1,1){\line(0,1){1}} \put(1,2){\line(1,0){1}} \put(2,1){\line(0,1){1}}\put(1,1){\makebox(1,1){\figuresize{$f_X$}}}
\put(1.50,0.80){\line(0,1){.3}}
\put(0,0){\line(1,0){1}} \put(0,0){\line(0,1){1}} \put(0,1){\line(1,0){1}} \put(1,0){\line(0,1){1}} \put(0,0){\line(1,1){1}} \put(1,0){\line(-1,1){1}}
\put(2,0){\line(1,0){1}} \put(2,0){\line(0,1){1}} \put(2,1){\line(1,0){1}} \put(3,0){\line(0,1){1}} \put(2,0){\line(1,1){1}} \put(3,0){\line(-1,1){1}}
\end{picture}
}

\newcommand{\Ufig}{
\begin{picture}(2,3)
\put(0,1){\line(1,0){1}} \put(0,1){\line(0,1){1}} \put(0,2){\line(1,0){1}} \put(1,1){\line(0,1){1}} \put(0,1){\line(1,1){1}} \put(1,1){\line(-1,1){1}}
\put(1,1){\line(1,0){1}} \put(1,1){\line(0,1){1}} \put(1,2){\line(1,0){1}} \put(2,1){\line(0,1){1}}\put(1,1){\makebox(1,1){\figuresize{$f_X$}}}
\put(1.50,0.80){\line(0,1){.3}}
\end{picture}
}

\newcommand{\Tfig}{
\begin{picture}(3,3)
\put(1,2){\line(1,0){1}} \put(1,2){\line(0,1){1}} \put(1,3){\line(1,0){1}} \put(2,2){\line(0,1){1}} \put(1,2){\line(1,1){1}} \put(2,2){\line(-1,1){1}}
\put(1,1){\line(1,0){1}} \put(1,1){\line(0,1){1}} \put(1,2){\line(1,0){1}} \put(2,1){\line(0,1){1}}\put(1,1){\makebox(1,1){\figuresize{$f_X$}}}
\put(1.50,0.80){\line(0,1){.3}}
\put(0,0){\line(1,0){1}} \put(0,0){\line(0,1){1}} \put(0,1){\line(1,0){1}} \put(1,0){\line(0,1){1}} \put(0,0){\line(1,1){1}} \put(1,0){\line(-1,1){1}}
\end{picture}
}

\newcommand{\Sfig}{
\begin{picture}(3,3)
\put(0,1){\line(1,0){1}} \put(0,1){\line(0,1){1}} \put(0,2){\line(1,0){1}} \put(1,1){\line(0,1){1}} \put(0,1){\line(1,1){1}} \put(1,1){\line(-1,1){1}}
\put(1,1){\line(1,0){1}} \put(1,1){\line(0,1){1}} \put(1,2){\line(1,0){1}} \put(2,1){\line(0,1){1}}\put(1,1){\makebox(1,1){\figuresize{$f_X$}}}
\put(1.50,0.80){\line(0,1){.3}}
\put(2,0){\line(1,0){1}} \put(2,0){\line(0,1){1}} \put(2,1){\line(1,0){1}} \put(3,0){\line(0,1){1}} \put(2,0){\line(1,1){1}} \put(3,0){\line(-1,1){1}}
\end{picture}
}

\newcommand{\Ronefig}{
\begin{picture}(2,3)
\put(1,2){\line(1,0){1}} \put(1,2){\line(0,1){1}} \put(1,3){\line(1,0){1}} \put(2,2){\line(0,1){1}} \put(1,2){\line(1,1){1}} \put(2,2){\line(-1,1){1}}
\put(0,1){\line(1,0){1}} \put(0,1){\line(0,1){1}} \put(0,2){\line(1,0){1}} \put(1,1){\line(0,1){1}} \put(0,1){\line(1,1){1}} \put(1,1){\line(-1,1){1}}
\put(1,1){\line(1,0){1}} \put(1,1){\line(0,1){1}} \put(1,2){\line(1,0){1}} \put(2,1){\line(0,1){1}}\put(1,1){\makebox(1,1){\figuresize{$f_X$}}}
\put(1.50,0.80){\line(0,1){.3}}
\end{picture}
}

\newcommand{\Rtwofig}{
\begin{picture}(3,3)
\put(0,1){\line(1,0){1}} \put(0,1){\line(0,1){1}} \put(0,2){\line(1,0){1}} \put(1,1){\line(0,1){1}} \put(0,1){\line(1,1){1}} \put(1,1){\line(-1,1){1}}
\put(1,1){\line(1,0){1}} \put(1,1){\line(0,1){1}} \put(1,2){\line(1,0){1}} \put(2,1){\line(0,1){1}}\put(1,1){\makebox(1,1){\figuresize{$f_X$}}}
\put(1.50,0.80){\line(0,1){.3}}
\put(2,1){\line(1,0){1}} \put(2,1){\line(0,1){1}} \put(2,2){\line(1,0){1}} \put(3,1){\line(0,1){1}} \put(2,1){\line(1,1){1}} \put(3,1){\line(-1,1){1}}
\end{picture}
}

\newcommand{\GammaOne}{
\begin{picture}(4,2)
\put(1,1){\line(1,0){1}} \put(1,1){\line(0,1){1}} \put(1,2){\line(1,0){1}} \put(2,1){\line(0,1){1}}
\put(2,1){\line(1,0){1}} \put(2,1){\line(0,1){1}} \put(2,2){\line(1,0){1}} \put(3,1){\line(0,1){1}}
\put(3,1){\line(1,0){1}} \put(3,1){\line(0,1){1}} \put(3,2){\line(1,0){1}} \put(4,1){\line(0,1){1}}
\put(0,0){\line(1,0){1}} \put(0,0){\line(0,1){1}} \put(0,1){\line(1,0){1}} \put(1,0){\line(0,1){1}}
\put(1,0){\line(1,0){1}} \put(1,0){\line(0,1){1}} \put(1,1){\line(1,0){1}} \put(2,0){\line(0,1){1}}
\put(2,0){\line(1,0){1}} \put(2,0){\line(0,1){1}} \put(2,1){\line(1,0){1}} \put(3,0){\line(0,1){1}}\put(2,0){\makebox(1,1){\figuresize{$f_X$}}}
\put(2.50,-0.20){\line(0,1){.3}}
\put(3,0){\line(1,0){1}} \put(3,0){\line(0,1){1}} \put(3,1){\line(1,0){1}} \put(4,0){\line(0,1){1}}
\end{picture}
}

\newcommand{\GammaTwo}{
\begin{picture}(5,2)
\put(1,1){\line(1,0){1}} \put(1,1){\line(0,1){1}} \put(1,2){\line(1,0){1}} \put(2,1){\line(0,1){1}}
\put(2,1){\line(1,0){1}} \put(2,1){\line(0,1){1}} \put(2,2){\line(1,0){1}} \put(3,1){\line(0,1){1}}
\put(3,1){\line(1,0){1}} \put(3,1){\line(0,1){1}} \put(3,2){\line(1,0){1}} \put(4,1){\line(0,1){1}}
\put(0,0){\line(1,0){1}} \put(0,0){\line(0,1){1}} \put(0,1){\line(1,0){1}} \put(1,0){\line(0,1){1}} \put(0,0){\line(1,1){1}} \put(1,0){\line(-1,1){1}}
\put(1,0){\line(1,0){1}} \put(1,0){\line(0,1){1}} \put(1,1){\line(1,0){1}} \put(2,0){\line(0,1){1}}
\put(2,0){\line(1,0){1}} \put(2,0){\line(0,1){1}} \put(2,1){\line(1,0){1}} \put(3,0){\line(0,1){1}}\put(2,0){\makebox(1,1){\figuresize{$f_X$}}}
\put(2.50,-0.20){\line(0,1){.3}}
\put(3,0){\line(1,0){1}} \put(3,0){\line(0,1){1}} \put(3,1){\line(1,0){1}} \put(4,0){\line(0,1){1}}
\put(4,0){\line(1,0){1}} \put(4,0){\line(0,1){1}} \put(4,1){\line(1,0){1}} \put(5,0){\line(0,1){1}} \put(4,0){\line(1,1){1}} \put(5,0){\line(-1,1){1}}
\end{picture}
}

\newcommand{\GammaThree}{
\begin{picture}(3,2)
\put(0,1){\line(1,0){1}} \put(0,1){\line(0,1){1}} \put(0,2){\line(1,0){1}} \put(1,1){\line(0,1){1}}
\put(1,1){\line(1,0){1}} \put(1,1){\line(0,1){1}} \put(1,2){\line(1,0){1}} \put(2,1){\line(0,1){1}}
\put(2,1){\line(1,0){1}} \put(2,1){\line(0,1){1}} \put(2,2){\line(1,0){1}} \put(3,1){\line(0,1){1}} \put(2,1){\line(1,1){1}} \put(3,1){\line(-1,1){1}}
\put(0,0){\line(1,0){1}} \put(0,0){\line(0,1){1}} \put(0,1){\line(1,0){1}} \put(1,0){\line(0,1){1}}
\put(1,0){\line(1,0){1}} \put(1,0){\line(0,1){1}} \put(1,1){\line(1,0){1}} \put(2,0){\line(0,1){1}}\put(1,0){\makebox(1,1){\figuresize{$f_X$}}}
\put(1.50,-0.20){\line(0,1){.3}}
\put(2,0){\line(1,0){1}} \put(2,0){\line(0,1){1}} \put(2,1){\line(1,0){1}} \put(3,0){\line(0,1){1}}
\end{picture}
}

\newcommand{\GammaFour}{
\begin{picture}(3,2)
\put(0,1){\line(1,0){1}} \put(0,1){\line(0,1){1}} \put(0,2){\line(1,0){1}} \put(1,1){\line(0,1){1}} \put(0,1){\line(1,1){1}} \put(1,1){\line(-1,1){1}}
\put(1,1){\line(1,0){1}} \put(1,1){\line(0,1){1}} \put(1,2){\line(1,0){1}} \put(2,1){\line(0,1){1}}
\put(2,1){\line(1,0){1}} \put(2,1){\line(0,1){1}} \put(2,2){\line(1,0){1}} \put(3,1){\line(0,1){1}} \put(2,1){\line(1,1){1}} \put(3,1){\line(-1,1){1}}
\put(0,0){\line(1,0){1}} \put(0,0){\line(0,1){1}} \put(0,1){\line(1,0){1}} \put(1,0){\line(0,1){1}}
\put(1,0){\line(1,0){1}} \put(1,0){\line(0,1){1}} \put(1,1){\line(1,0){1}} \put(2,0){\line(0,1){1}}\put(1,0){\makebox(1,1){\figuresize{$f_X$}}}
\put(1.50,-0.20){\line(0,1){.3}}
\put(2,0){\line(1,0){1}} \put(2,0){\line(0,1){1}} \put(2,1){\line(1,0){1}} \put(3,0){\line(0,1){1}}
\end{picture}
}

\newcommand{\GammaFive}{
\begin{picture}(2,1)
\put(0,0){\line(1,0){1}} \put(0,0){\line(0,1){1}} \put(0,1){\line(1,0){1}} \put(1,0){\line(0,1){1}} \put(0,0){\line(1,1){1}} \put(1,0){\line(-1,1){1}}
\put(1,0){\line(1,0){1}} \put(1,0){\line(0,1){1}} \put(1,1){\line(1,0){1}} \put(2,0){\line(0,1){1}}\put(1,0){\makebox(1,1){\figuresize{$f_X$}}}
\put(1.50,-0.20){\line(0,1){.3}}
\end{picture}
}

\newcommand{\GammaSix}{
\begin{picture}(3,2)
\put(1,1){\line(1,0){1}} \put(1,1){\line(0,1){1}} \put(1,2){\line(1,0){1}} \put(2,1){\line(0,1){1}} \put(1,1){\line(1,1){1}} \put(2,1){\line(-1,1){1}}
\put(0,0){\line(1,0){1}} \put(0,0){\line(0,1){1}} \put(0,1){\line(1,0){1}} \put(1,0){\line(0,1){1}}
\put(1,0){\line(1,0){1}} \put(1,0){\line(0,1){1}} \put(1,1){\line(1,0){1}} \put(2,0){\line(0,1){1}}\put(1,0){\makebox(1,1){\figuresize{$f_X$}}}
\put(1.50,-0.20){\line(0,1){.3}}
\put(2,0){\line(1,0){1}} \put(2,0){\line(0,1){1}} \put(2,1){\line(1,0){1}} \put(3,0){\line(0,1){1}}
\end{picture}
}

\newcommand{\VOneA}{
\begin{picture}(4,3)
\put(1,2){\line(1,0){1}} \put(1,2){\line(0,1){1}} \put(1,3){\line(1,0){1}} \put(2,2){\line(0,1){1}}
\put(2,2){\line(1,0){1}} \put(2,2){\line(0,1){1}} \put(2,3){\line(1,0){1}} \put(3,2){\line(0,1){1}}
\put(3,2){\line(1,0){1}} \put(3,2){\line(0,1){1}} \put(3,3){\line(1,0){1}} \put(4,2){\line(0,1){1}}
\put(0,1){\line(1,0){1}} \put(0,1){\line(0,1){1}} \put(0,2){\line(1,0){1}} \put(1,1){\line(0,1){1}}
\put(1,1){\line(1,0){1}} \put(1,1){\line(0,1){1}} \put(1,2){\line(1,0){1}} \put(2,1){\line(0,1){1}}
\put(2,1){\line(1,0){1}} \put(2,1){\line(0,1){1}} \put(2,2){\line(1,0){1}} \put(3,1){\line(0,1){1}}\put(2,1){\makebox(1,1){\figuresize{$f_X$}}}
\put(2.50,0.80){\line(0,1){.3}}
\put(3,1){\line(1,0){1}} \put(3,1){\line(0,1){1}} \put(3,2){\line(1,0){1}} \put(4,1){\line(0,1){1}}
\put(1,0){\line(1,0){1}} \put(1,0){\line(0,1){1}} \put(1,1){\line(1,0){1}} \put(2,0){\line(0,1){1}} \put(1,0){\line(1,1){1}} \put(2,0){\line(-1,1){1}}
\end{picture}
}

\newcommand{\VOneB}{
\begin{picture}(4,3)
\put(1,2){\line(1,0){1}} \put(1,2){\line(0,1){1}} \put(1,3){\line(1,0){1}} \put(2,2){\line(0,1){1}}
\put(2,2){\line(1,0){1}} \put(2,2){\line(0,1){1}} \put(2,3){\line(1,0){1}} \put(3,2){\line(0,1){1}}
\put(3,2){\line(1,0){1}} \put(3,2){\line(0,1){1}} \put(3,3){\line(1,0){1}} \put(4,2){\line(0,1){1}}
\put(0,1){\line(1,0){1}} \put(0,1){\line(0,1){1}} \put(0,2){\line(1,0){1}} \put(1,1){\line(0,1){1}}
\put(1,1){\line(1,0){1}} \put(1,1){\line(0,1){1}} \put(1,2){\line(1,0){1}} \put(2,1){\line(0,1){1}}
\put(2,1){\line(1,0){1}} \put(2,1){\line(0,1){1}} \put(2,2){\line(1,0){1}} \put(3,1){\line(0,1){1}}\put(2,1){\makebox(1,1){\figuresize{$f_X$}}}
\put(2.50,0.80){\line(0,1){.3}}
\put(3,1){\line(1,0){1}} \put(3,1){\line(0,1){1}} \put(3,2){\line(1,0){1}} \put(4,1){\line(0,1){1}}
\put(3,0){\line(1,0){1}} \put(3,0){\line(0,1){1}} \put(3,1){\line(1,0){1}} \put(4,0){\line(0,1){1}} \put(3,0){\line(1,1){1}} \put(4,0){\line(-1,1){1}}
\end{picture}
}

\newcommand{\VTwo}{
\begin{picture}(5,3)
\put(1,2){\line(1,0){1}} \put(1,2){\line(0,1){1}} \put(1,3){\line(1,0){1}} \put(2,2){\line(0,1){1}}
\put(2,2){\line(1,0){1}} \put(2,2){\line(0,1){1}} \put(2,3){\line(1,0){1}} \put(3,2){\line(0,1){1}}
\put(3,2){\line(1,0){1}} \put(3,2){\line(0,1){1}} \put(3,3){\line(1,0){1}} \put(4,2){\line(0,1){1}}
\put(0,1){\line(1,0){1}} \put(0,1){\line(0,1){1}} \put(0,2){\line(1,0){1}} \put(1,1){\line(0,1){1}} \put(0,1){\line(1,1){1}} \put(1,1){\line(-1,1){1}}
\put(1,1){\line(1,0){1}} \put(1,1){\line(0,1){1}} \put(1,2){\line(1,0){1}} \put(2,1){\line(0,1){1}}
\put(2,1){\line(1,0){1}} \put(2,1){\line(0,1){1}} \put(2,2){\line(1,0){1}} \put(3,1){\line(0,1){1}}\put(2,1){\makebox(1,1){\figuresize{$f_X$}}}
\put(2.50,0.80){\line(0,1){.3}}
\put(3,1){\line(1,0){1}} \put(3,1){\line(0,1){1}} \put(3,2){\line(1,0){1}} \put(4,1){\line(0,1){1}}
\put(4,1){\line(1,0){1}} \put(4,1){\line(0,1){1}} \put(4,2){\line(1,0){1}} \put(5,1){\line(0,1){1}} \put(4,1){\line(1,1){1}} \put(5,1){\line(-1,1){1}}
\put(1,0){\line(1,0){1}} \put(1,0){\line(0,1){1}} \put(1,1){\line(1,0){1}} \put(2,0){\line(0,1){1}} \put(1,0){\line(1,1){1}} \put(2,0){\line(-1,1){1}}
\end{picture}
}

\newcommand{\VThree}{
\begin{picture}(3,3)
\put(0,2){\line(1,0){1}} \put(0,2){\line(0,1){1}} \put(0,3){\line(1,0){1}} \put(1,2){\line(0,1){1}}
\put(1,2){\line(1,0){1}} \put(1,2){\line(0,1){1}} \put(1,3){\line(1,0){1}} \put(2,2){\line(0,1){1}}
\put(2,2){\line(1,0){1}} \put(2,2){\line(0,1){1}} \put(2,3){\line(1,0){1}} \put(3,2){\line(0,1){1}} \put(2,2){\line(1,1){1}} \put(3,2){\line(-1,1){1}}
\put(0,1){\line(1,0){1}} \put(0,1){\line(0,1){1}} \put(0,2){\line(1,0){1}} \put(1,1){\line(0,1){1}}
\put(1,1){\line(1,0){1}} \put(1,1){\line(0,1){1}} \put(1,2){\line(1,0){1}} \put(2,1){\line(0,1){1}}\put(1,1){\makebox(1,1){\figuresize{$f_X$}}}
\put(1.50,0.80){\line(0,1){.3}}
\put(2,1){\line(1,0){1}} \put(2,1){\line(0,1){1}} \put(2,2){\line(1,0){1}} \put(3,1){\line(0,1){1}}
\put(0,0){\line(1,0){1}} \put(0,0){\line(0,1){1}} \put(0,1){\line(1,0){1}} \put(1,0){\line(0,1){1}} \put(0,0){\line(1,1){1}} \put(1,0){\line(-1,1){1}}
\end{picture}
}

\newcommand{\VFour}{
\begin{picture}(3,3)
\put(0,2){\line(1,0){1}} \put(0,2){\line(0,1){1}} \put(0,3){\line(1,0){1}} \put(1,2){\line(0,1){1}} \put(0,2){\line(1,1){1}} \put(1,2){\line(-1,1){1}}
\put(1,2){\line(1,0){1}} \put(1,2){\line(0,1){1}} \put(1,3){\line(1,0){1}} \put(2,2){\line(0,1){1}}
\put(2,2){\line(1,0){1}} \put(2,2){\line(0,1){1}} \put(2,3){\line(1,0){1}} \put(3,2){\line(0,1){1}} \put(2,2){\line(1,1){1}} \put(3,2){\line(-1,1){1}}
\put(0,1){\line(1,0){1}} \put(0,1){\line(0,1){1}} \put(0,2){\line(1,0){1}} \put(1,1){\line(0,1){1}}
\put(1,1){\line(1,0){1}} \put(1,1){\line(0,1){1}} \put(1,2){\line(1,0){1}} \put(2,1){\line(0,1){1}}\put(1,1){\makebox(1,1){\figuresize{$f_X$}}}
\put(1.50,0.80){\line(0,1){.3}}
\put(2,1){\line(1,0){1}} \put(2,1){\line(0,1){1}} \put(2,2){\line(1,0){1}} \put(3,1){\line(0,1){1}}
\put(0,0){\line(1,0){1}} \put(0,0){\line(0,1){1}} \put(0,1){\line(1,0){1}} \put(1,0){\line(0,1){1}} \put(0,0){\line(1,1){1}} \put(1,0){\line(-1,1){1}}
\end{picture}
}

\newcommand{\VFiveA}{
\begin{picture}(2,2)
\put(0,1){\line(1,0){1}} \put(0,1){\line(0,1){1}} \put(0,2){\line(1,0){1}} \put(1,1){\line(0,1){1}} \put(0,1){\line(1,1){1}} \put(1,1){\line(-1,1){1}}
\put(1,1){\line(1,0){1}} \put(1,1){\line(0,1){1}} \put(1,2){\line(1,0){1}} \put(2,1){\line(0,1){1}}\put(1,1){\makebox(1,1){\figuresize{$f_X$}}}
\put(1.50,0.80){\line(0,1){.3}}
\put(0,0){\line(1,0){1}} \put(0,0){\line(0,1){1}} \put(0,1){\line(1,0){1}} \put(1,0){\line(0,1){1}} \put(0,0){\line(1,1){1}} \put(1,0){\line(-1,1){1}}
\end{picture}
}

\newcommand{\VFiveB}{
\begin{picture}(3,2)
\put(0,1){\line(1,0){1}} \put(0,1){\line(0,1){1}} \put(0,2){\line(1,0){1}} \put(1,1){\line(0,1){1}} \put(0,1){\line(1,1){1}} \put(1,1){\line(-1,1){1}}
\put(1,1){\line(1,0){1}} \put(1,1){\line(0,1){1}} \put(1,2){\line(1,0){1}} \put(2,1){\line(0,1){1}}\put(1,1){\makebox(1,1){\figuresize{$f_X$}}}
\put(1.50,0.80){\line(0,1){.3}}
\put(2,0){\line(1,0){1}} \put(2,0){\line(0,1){1}} \put(2,1){\line(1,0){1}} \put(3,0){\line(0,1){1}} \put(2,0){\line(1,1){1}} \put(3,0){\line(-1,1){1}}
\end{picture}
}

\newcommand{\VSix}{
\begin{picture}(3,3)
\put(1,2){\line(1,0){1}} \put(1,2){\line(0,1){1}} \put(1,3){\line(1,0){1}} \put(2,2){\line(0,1){1}} \put(1,2){\line(1,1){1}} \put(2,2){\line(-1,1){1}}
\put(0,1){\line(1,0){1}} \put(0,1){\line(0,1){1}} \put(0,2){\line(1,0){1}} \put(1,1){\line(0,1){1}}
\put(1,1){\line(1,0){1}} \put(1,1){\line(0,1){1}} \put(1,2){\line(1,0){1}} \put(2,1){\line(0,1){1}}\put(1,1){\makebox(1,1){\figuresize{$f_X$}}}
\put(1.50,0.80){\line(0,1){.3}}
\put(2,1){\line(1,0){1}} \put(2,1){\line(0,1){1}} \put(2,2){\line(1,0){1}} \put(3,1){\line(0,1){1}}
\put(0,0){\line(1,0){1}} \put(0,0){\line(0,1){1}} \put(0,1){\line(1,0){1}} \put(1,0){\line(0,1){1}} \put(0,0){\line(1,1){1}} \put(1,0){\line(-1,1){1}}
\end{picture}
}

\newcommand{\VSeven}{
\begin{picture}(3,3)
\put(0,2){\line(1,0){1}} \put(0,2){\line(0,1){1}} \put(0,3){\line(1,0){1}} \put(1,2){\line(0,1){1}}
\put(1,2){\line(1,0){1}} \put(1,2){\line(0,1){1}} \put(1,3){\line(1,0){1}} \put(2,2){\line(0,1){1}}
\put(2,2){\line(1,0){1}} \put(2,2){\line(0,1){1}} \put(2,3){\line(1,0){1}} \put(3,2){\line(0,1){1}} \put(2,2){\line(1,1){1}} \put(3,2){\line(-1,1){1}}
\put(0,1){\line(1,0){1}} \put(0,1){\line(0,1){1}} \put(0,2){\line(1,0){1}} \put(1,1){\line(0,1){1}}
\put(1,1){\line(1,0){1}} \put(1,1){\line(0,1){1}} \put(1,2){\line(1,0){1}} \put(2,1){\line(0,1){1}}\put(1,1){\makebox(1,1){\figuresize{$f_X$}}}
\put(1.50,0.80){\line(0,1){.3}}
\put(2,1){\line(1,0){1}} \put(2,1){\line(0,1){1}} \put(2,2){\line(1,0){1}} \put(3,1){\line(0,1){1}}
\put(2,0){\line(1,0){1}} \put(2,0){\line(0,1){1}} \put(2,1){\line(1,0){1}} \put(3,0){\line(0,1){1}} \put(2,0){\line(1,1){1}} \put(3,0){\line(-1,1){1}}
\end{picture}
}

\newcommand{\WOne}{
\begin{picture}(3,3)
\put(1,2){\line(1,0){1}} \put(1,2){\line(0,1){1}} \put(1,3){\line(1,0){1}} \put(2,2){\line(0,1){1}}
\put(0,1){\line(1,0){1}} \put(0,1){\line(0,1){1}} \put(0,2){\line(1,0){1}} \put(1,1){\line(0,1){1}}
\put(1,1){\line(1,0){1}} \put(1,1){\line(0,1){1}} \put(1,2){\line(1,0){1}} \put(2,1){\line(0,1){1}}\put(1,1){\makebox(1,1){\figuresize{$f_X$}}}
\put(1.50,0.80){\line(0,1){.3}}
\put(2,1){\line(1,0){1}} \put(2,1){\line(0,1){1}} \put(2,2){\line(1,0){1}} \put(3,1){\line(0,1){1}}
\put(0,0){\line(1,0){1}} \put(0,0){\line(0,1){1}} \put(0,1){\line(1,0){1}} \put(1,0){\line(0,1){1}} \put(0,0){\line(1,1){1}} \put(1,0){\line(-1,1){1}}
\put(2,0){\line(1,0){1}} \put(2,0){\line(0,1){1}} \put(2,1){\line(1,0){1}} \put(3,0){\line(0,1){1}} \put(2,0){\line(1,1){1}} \put(3,0){\line(-1,1){1}}
\end{picture}
}

\newcommand{\WTwo}{
\begin{picture}(3,2)
\put(0,1){\line(1,0){1}} \put(0,1){\line(0,1){1}} \put(0,2){\line(1,0){1}} \put(1,1){\line(0,1){1}} \put(0,1){\line(1,1){1}} \put(1,1){\line(-1,1){1}}
\put(1,1){\line(1,0){1}} \put(1,1){\line(0,1){1}} \put(1,2){\line(1,0){1}} \put(2,1){\line(0,1){1}}\put(1,1){\makebox(1,1){\figuresize{$f_X$}}}
\put(1.50,0.80){\line(0,1){.3}}
\put(0,0){\line(1,0){1}} \put(0,0){\line(0,1){1}} \put(0,1){\line(1,0){1}} \put(1,0){\line(0,1){1}} \put(0,0){\line(1,1){1}} \put(1,0){\line(-1,1){1}}
\put(2,0){\line(1,0){1}} \put(2,0){\line(0,1){1}} \put(2,1){\line(1,0){1}} \put(3,0){\line(0,1){1}} \put(2,0){\line(1,1){1}} \put(3,0){\line(-1,1){1}}
\end{picture}
}

\newcommand{\WThree}{
\begin{picture}(3,3)
\put(1,2){\line(1,0){1}} \put(1,2){\line(0,1){1}} \put(1,3){\line(1,0){1}} \put(2,2){\line(0,1){1}} \put(1,2){\line(1,1){1}} \put(2,2){\line(-1,1){1}}
\put(0,1){\line(1,0){1}} \put(0,1){\line(0,1){1}} \put(0,2){\line(1,0){1}} \put(1,1){\line(0,1){1}}
\put(1,1){\line(1,0){1}} \put(1,1){\line(0,1){1}} \put(1,2){\line(1,0){1}} \put(2,1){\line(0,1){1}}\put(1,1){\makebox(1,1){\figuresize{$f_X$}}}
\put(1.50,0.80){\line(0,1){.3}}
\put(2,1){\line(1,0){1}} \put(2,1){\line(0,1){1}} \put(2,2){\line(1,0){1}} \put(3,1){\line(0,1){1}}
\put(0,0){\line(1,0){1}} \put(0,0){\line(0,1){1}} \put(0,1){\line(1,0){1}} \put(1,0){\line(0,1){1}} \put(0,0){\line(1,1){1}} \put(1,0){\line(-1,1){1}}
\put(2,0){\line(1,0){1}} \put(2,0){\line(0,1){1}} \put(2,1){\line(1,0){1}} \put(3,0){\line(0,1){1}} \put(2,0){\line(1,1){1}} \put(3,0){\line(-1,1){1}}
\end{picture}
}

\newcommand{\UOne}{
\begin{picture}(3,2)
\put(0,1){\line(1,0){1}} \put(0,1){\line(0,1){1}} \put(0,2){\line(1,0){1}} \put(1,1){\line(0,1){1}}
\put(1,1){\line(1,0){1}} \put(1,1){\line(0,1){1}} \put(1,2){\line(1,0){1}} \put(2,1){\line(0,1){1}}
\put(2,1){\line(1,0){1}} \put(2,1){\line(0,1){1}} \put(2,2){\line(1,0){1}} \put(3,1){\line(0,1){1}}
\put(0,0){\line(1,0){1}} \put(0,0){\line(0,1){1}} \put(0,1){\line(1,0){1}} \put(1,0){\line(0,1){1}}
\put(1,0){\line(1,0){1}} \put(1,0){\line(0,1){1}} \put(1,1){\line(1,0){1}} \put(2,0){\line(0,1){1}}\put(1,0){\makebox(1,1){\figuresize{$f_X$}}}
\put(1.50,-0.20){\line(0,1){.3}}
\put(2,0){\line(1,0){1}} \put(2,0){\line(0,1){1}} \put(2,1){\line(1,0){1}} \put(3,0){\line(0,1){1}} \put(2,0){\line(1,1){1}} \put(3,0){\line(-1,1){1}}
\end{picture}
}

\newcommand{\UTwo}{
\begin{picture}(3,2)
\put(0,1){\line(1,0){1}} \put(0,1){\line(0,1){1}} \put(0,2){\line(1,0){1}} \put(1,1){\line(0,1){1}}
\put(1,1){\line(1,0){1}} \put(1,1){\line(0,1){1}} \put(1,2){\line(1,0){1}} \put(2,1){\line(0,1){1}}
\put(2,1){\line(1,0){1}} \put(2,1){\line(0,1){1}} \put(2,2){\line(1,0){1}} \put(3,1){\line(0,1){1}} \put(2,1){\line(1,1){1}} \put(3,1){\line(-1,1){1}}
\put(0,0){\line(1,0){1}} \put(0,0){\line(0,1){1}} \put(0,1){\line(1,0){1}} \put(1,0){\line(0,1){1}}
\put(1,0){\line(1,0){1}} \put(1,0){\line(0,1){1}} \put(1,1){\line(1,0){1}} \put(2,0){\line(0,1){1}}\put(1,0){\makebox(1,1){\figuresize{$f_X$}}}
\put(1.50,-0.20){\line(0,1){.3}}
\put(2,0){\line(1,0){1}} \put(2,0){\line(0,1){1}} \put(2,1){\line(1,0){1}} \put(3,0){\line(0,1){1}} \put(2,0){\line(1,1){1}} \put(3,0){\line(-1,1){1}}
\end{picture}
}

\newcommand{\UThree}{
\begin{picture}(3,2)
\put(0,1){\line(1,0){1}} \put(0,1){\line(0,1){1}} \put(0,2){\line(1,0){1}} \put(1,1){\line(0,1){1}} \put(0,1){\line(1,1){1}} \put(1,1){\line(-1,1){1}}
\put(1,1){\line(1,0){1}} \put(1,1){\line(0,1){1}} \put(1,2){\line(1,0){1}} \put(2,1){\line(0,1){1}}
\put(0,0){\line(1,0){1}} \put(0,0){\line(0,1){1}} \put(0,1){\line(1,0){1}} \put(1,0){\line(0,1){1}}
\put(1,0){\line(1,0){1}} \put(1,0){\line(0,1){1}} \put(1,1){\line(1,0){1}} \put(2,0){\line(0,1){1}}\put(1,0){\makebox(1,1){\figuresize{$f_X$}}}
\put(1.50,-0.20){\line(0,1){.3}}
\put(2,0){\line(1,0){1}} \put(2,0){\line(0,1){1}} \put(2,1){\line(1,0){1}} \put(3,0){\line(0,1){1}} \put(2,0){\line(1,1){1}} \put(3,0){\line(-1,1){1}}
\end{picture}
}

\newcommand{\UFourA}{
\begin{picture}(2,2)
\put(0,1){\line(1,0){1}} \put(0,1){\line(0,1){1}} \put(0,2){\line(1,0){1}} \put(1,1){\line(0,1){1}} \put(0,1){\line(1,1){1}} \put(1,1){\line(-1,1){1}}
\put(0,0){\line(1,0){1}} \put(0,0){\line(0,1){1}} \put(0,1){\line(1,0){1}} \put(1,0){\line(0,1){1}}\put(0,0){\makebox(1,1){\figuresize{$f_X$}}}
\put(0.50,-0.20){\line(0,1){.3}}
\put(1,0){\line(1,0){1}} \put(1,0){\line(0,1){1}} \put(1,1){\line(1,0){1}} \put(2,0){\line(0,1){1}} \put(1,0){\line(1,1){1}} \put(2,0){\line(-1,1){1}}
\end{picture}
}

\newcommand{\UFourB}{
\begin{picture}(3,1)
\put(0,0){\line(1,0){1}} \put(0,0){\line(0,1){1}} \put(0,1){\line(1,0){1}} \put(1,0){\line(0,1){1}} \put(0,0){\line(1,1){1}} \put(1,0){\line(-1,1){1}}
\put(1,0){\line(1,0){1}} \put(1,0){\line(0,1){1}} \put(1,1){\line(1,0){1}} \put(2,0){\line(0,1){1}}\put(1,0){\makebox(1,1){\figuresize{$f_X$}}}
\put(1.50,-0.20){\line(0,1){.3}}
\put(2,0){\line(1,0){1}} \put(2,0){\line(0,1){1}} \put(2,1){\line(1,0){1}} \put(3,0){\line(0,1){1}} \put(2,0){\line(1,1){1}} \put(3,0){\line(-1,1){1}}
\end{picture}
}

\newcommand{\SOneA}{
\begin{picture}(3,2)
\put(0,1){\line(1,0){1}} \put(0,1){\line(0,1){1}} \put(0,2){\line(1,0){1}} \put(1,1){\line(0,1){1}} \put(0,1){\line(1,1){1}} \put(1,1){\line(-1,1){1}}
\put(1,1){\line(1,0){1}} \put(1,1){\line(0,1){1}} \put(1,2){\line(1,0){1}} \put(2,1){\line(0,1){1}}\put(1,1){\makebox(1,1){\figuresize{$f_X$}}}
\put(1.50,0.80){\line(0,1){.3}}
\put(2,1){\line(1,0){1}} \put(2,1){\line(0,1){1}} \put(2,2){\line(1,0){1}} \put(3,1){\line(0,1){1}} \put(2,1){\line(1,1){1}} \put(3,1){\line(-1,1){1}}
\put(0,0){\line(1,0){1}} \put(0,0){\line(0,1){1}} \put(0,1){\line(1,0){1}} \put(1,0){\line(0,1){1}} \put(0,0){\line(1,1){1}} \put(1,0){\line(-1,1){1}}
\end{picture}
}

\newcommand{\SOneB}{
\begin{picture}(3,3)
\put(1,2){\line(1,0){1}} \put(1,2){\line(0,1){1}} \put(1,3){\line(1,0){1}} \put(2,2){\line(0,1){1}} \put(1,2){\line(1,1){1}} \put(2,2){\line(-1,1){1}}
\put(1,1){\line(1,0){1}} \put(1,1){\line(0,1){1}} \put(1,2){\line(1,0){1}} \put(2,1){\line(0,1){1}}\put(1,1){\makebox(1,1){\figuresize{$f_X$}}}
\put(1.50,0.80){\line(0,1){.3}}
\put(2,1){\line(1,0){1}} \put(2,1){\line(0,1){1}} \put(2,2){\line(1,0){1}} \put(3,1){\line(0,1){1}} \put(2,1){\line(1,1){1}} \put(3,1){\line(-1,1){1}}
\put(0,0){\line(1,0){1}} \put(0,0){\line(0,1){1}} \put(0,1){\line(1,0){1}} \put(1,0){\line(0,1){1}} \put(0,0){\line(1,1){1}} \put(1,0){\line(-1,1){1}}
\end{picture}
}

\newcommand{\STwo}{
\begin{picture}(3,3)
\put(0,2){\line(1,0){1}} \put(0,2){\line(0,1){1}} \put(0,3){\line(1,0){1}} \put(1,2){\line(0,1){1}}
\put(1,2){\line(1,0){1}} \put(1,2){\line(0,1){1}} \put(1,3){\line(1,0){1}} \put(2,2){\line(0,1){1}}
\put(0,1){\line(1,0){1}} \put(0,1){\line(0,1){1}} \put(0,2){\line(1,0){1}} \put(1,1){\line(0,1){1}}
\put(1,1){\line(1,0){1}} \put(1,1){\line(0,1){1}} \put(1,2){\line(1,0){1}} \put(2,1){\line(0,1){1}}\put(1,1){\makebox(1,1){\figuresize{$f_X$}}}
\put(1.50,0.80){\line(0,1){.3}}
\put(2,1){\line(1,0){1}} \put(2,1){\line(0,1){1}} \put(2,2){\line(1,0){1}} \put(3,1){\line(0,1){1}} \put(2,1){\line(1,1){1}} \put(3,1){\line(-1,1){1}}
\put(0,0){\line(1,0){1}} \put(0,0){\line(0,1){1}} \put(0,1){\line(1,0){1}} \put(1,0){\line(0,1){1}} \put(0,0){\line(1,1){1}} \put(1,0){\line(-1,1){1}}
\end{picture}
}

\newcommand{\SThree}{
\begin{picture}(3,3)
\put(0,2){\line(1,0){1}} \put(0,2){\line(0,1){1}} \put(0,3){\line(1,0){1}} \put(1,2){\line(0,1){1}} \put(0,2){\line(1,1){1}} \put(1,2){\line(-1,1){1}}
\put(1,2){\line(1,0){1}} \put(1,2){\line(0,1){1}} \put(1,3){\line(1,0){1}} \put(2,2){\line(0,1){1}}
\put(0,1){\line(1,0){1}} \put(0,1){\line(0,1){1}} \put(0,2){\line(1,0){1}} \put(1,1){\line(0,1){1}}
\put(1,1){\line(1,0){1}} \put(1,1){\line(0,1){1}} \put(1,2){\line(1,0){1}} \put(2,1){\line(0,1){1}}\put(1,1){\makebox(1,1){\figuresize{$f_X$}}}
\put(1.50,0.80){\line(0,1){.3}}
\put(2,1){\line(1,0){1}} \put(2,1){\line(0,1){1}} \put(2,2){\line(1,0){1}} \put(3,1){\line(0,1){1}} \put(2,1){\line(1,1){1}} \put(3,1){\line(-1,1){1}}
\put(0,0){\line(1,0){1}} \put(0,0){\line(0,1){1}} \put(0,1){\line(1,0){1}} \put(1,0){\line(0,1){1}} \put(0,0){\line(1,1){1}} \put(1,0){\line(-1,1){1}}
\end{picture}
}

\newcommand{\TOne}{
\begin{picture}(3,3)
\put(1,2){\line(1,0){1}} \put(1,2){\line(0,1){1}} \put(1,3){\line(1,0){1}} \put(2,2){\line(0,1){1}} \put(1,2){\line(1,1){1}} \put(2,2){\line(-1,1){1}}
\put(0,1){\line(1,0){1}} \put(0,1){\line(0,1){1}} \put(0,2){\line(1,0){1}} \put(1,1){\line(0,1){1}}
\put(1,1){\line(1,0){1}} \put(1,1){\line(0,1){1}} \put(1,2){\line(1,0){1}} \put(2,1){\line(0,1){1}}\put(1,1){\makebox(1,1){\figuresize{$f_X$}}}
\put(1.50,0.80){\line(0,1){.3}}
\put(2,1){\line(1,0){1}} \put(2,1){\line(0,1){1}} \put(2,2){\line(1,0){1}} \put(3,1){\line(0,1){1}}
\put(2,0){\line(1,0){1}} \put(2,0){\line(0,1){1}} \put(2,1){\line(1,0){1}} \put(3,0){\line(0,1){1}} \put(2,0){\line(1,1){1}} \put(3,0){\line(-1,1){1}}
\end{picture}
}

\newcommand{\TTwo}{
\begin{picture}(3,3)
\put(1,2){\line(1,0){1}} \put(1,2){\line(0,1){1}} \put(1,3){\line(1,0){1}} \put(2,2){\line(0,1){1}} \put(1,2){\line(1,1){1}} \put(2,2){\line(-1,1){1}}
\put(0,1){\line(1,0){1}} \put(0,1){\line(0,1){1}} \put(0,2){\line(1,0){1}} \put(1,1){\line(0,1){1}}
\put(1,1){\line(1,0){1}} \put(1,1){\line(0,1){1}} \put(1,2){\line(1,0){1}} \put(2,1){\line(0,1){1}}\put(1,1){\makebox(1,1){\figuresize{$f_X$}}}
\put(1.50,0.80){\line(0,1){.3}}
\put(2,1){\line(1,0){1}} \put(2,1){\line(0,1){1}} \put(2,2){\line(1,0){1}} \put(3,1){\line(0,1){1}} \put(2,1){\line(1,1){1}} \put(3,1){\line(-1,1){1}}
\put(2,0){\line(1,0){1}} \put(2,0){\line(0,1){1}} \put(2,1){\line(1,0){1}} \put(3,0){\line(0,1){1}} \put(2,0){\line(1,1){1}} \put(3,0){\line(-1,1){1}}
\end{picture}
}

\newcommand{\TThree}{
\begin{picture}(3,3)
\put(1,2){\line(1,0){1}} \put(1,2){\line(0,1){1}} \put(1,3){\line(1,0){1}} \put(2,2){\line(0,1){1}} \put(1,2){\line(1,1){1}} \put(2,2){\line(-1,1){1}}
\put(0,1){\line(1,0){1}} \put(0,1){\line(0,1){1}} \put(0,2){\line(1,0){1}} \put(1,1){\line(0,1){1}} \put(0,1){\line(1,1){1}} \put(1,1){\line(-1,1){1}}
\put(1,1){\line(1,0){1}} \put(1,1){\line(0,1){1}} \put(1,2){\line(1,0){1}} \put(2,1){\line(0,1){1}}\put(1,1){\makebox(1,1){\figuresize{$f_X$}}}
\put(1.50,0.80){\line(0,1){.3}}
\put(2,1){\line(1,0){1}} \put(2,1){\line(0,1){1}} \put(2,2){\line(1,0){1}} \put(3,1){\line(0,1){1}}
\put(2,0){\line(1,0){1}} \put(2,0){\line(0,1){1}} \put(2,1){\line(1,0){1}} \put(3,0){\line(0,1){1}} \put(2,0){\line(1,1){1}} \put(3,0){\line(-1,1){1}}
\end{picture}
}

\newcommand{\ROneA}{
\begin{picture}(3,2)
\put(1,1){\line(1,0){1}} \put(1,1){\line(0,1){1}} \put(1,2){\line(1,0){1}} \put(2,1){\line(0,1){1}} \put(1,1){\line(1,1){1}} \put(2,1){\line(-1,1){1}}
\put(0,0){\line(1,0){1}} \put(0,0){\line(0,1){1}} \put(0,1){\line(1,0){1}} \put(1,0){\line(0,1){1}}
\put(1,0){\line(1,0){1}} \put(1,0){\line(0,1){1}} \put(1,1){\line(1,0){1}} \put(2,0){\line(0,1){1}}\put(1,0){\makebox(1,1){\figuresize{$f_X$}}}
\put(1.50,-0.20){\line(0,1){.3}}
\put(2,0){\line(1,0){1}} \put(2,0){\line(0,1){1}} \put(2,1){\line(1,0){1}} \put(3,0){\line(0,1){1}} \put(2,0){\line(1,1){1}} \put(3,0){\line(-1,1){1}}
\end{picture}
}

\newcommand{\ROneB}{
\begin{picture}(3,2)
\put(1,1){\line(1,0){1}} \put(1,1){\line(0,1){1}} \put(1,2){\line(1,0){1}} \put(2,1){\line(0,1){1}}
\put(0,0){\line(1,0){1}} \put(0,0){\line(0,1){1}} \put(0,1){\line(1,0){1}} \put(1,0){\line(0,1){1}} \put(0,0){\line(1,1){1}} \put(1,0){\line(-1,1){1}}
\put(1,0){\line(1,0){1}} \put(1,0){\line(0,1){1}} \put(1,1){\line(1,0){1}} \put(2,0){\line(0,1){1}}\put(1,0){\makebox(1,1){\figuresize{$f_X$}}}
\put(1.50,-0.20){\line(0,1){.3}}
\put(2,0){\line(1,0){1}} \put(2,0){\line(0,1){1}} \put(2,1){\line(1,0){1}} \put(3,0){\line(0,1){1}} \put(2,0){\line(1,1){1}} \put(3,0){\line(-1,1){1}}
\end{picture}
}

\newcommand{\RTwo}{
\begin{picture}(3,2)
\put(1,1){\line(1,0){1}} \put(1,1){\line(0,1){1}} \put(1,2){\line(1,0){1}} \put(2,1){\line(0,1){1}} \put(1,1){\line(1,1){1}} \put(2,1){\line(-1,1){1}}
\put(0,0){\line(1,0){1}} \put(0,0){\line(0,1){1}} \put(0,1){\line(1,0){1}} \put(1,0){\line(0,1){1}} \put(0,0){\line(1,1){1}} \put(1,0){\line(-1,1){1}}
\put(1,0){\line(1,0){1}} \put(1,0){\line(0,1){1}} \put(1,1){\line(1,0){1}} \put(2,0){\line(0,1){1}}\put(1,0){\makebox(1,1){\figuresize{$f_X$}}}
\put(1.50,-0.20){\line(0,1){.3}}
\put(2,0){\line(1,0){1}} \put(2,0){\line(0,1){1}} \put(2,1){\line(1,0){1}} \put(3,0){\line(0,1){1}} \put(2,0){\line(1,1){1}} \put(3,0){\line(-1,1){1}}
\end{picture}
}

\newcommand{\figuresize}{\scriptsize}
\newcommand{\caseConvex}[2]{\parbox[b][2.7cm][b]{3.5cm}{\center #1 \\ \vspace{1mm} #2 \\ \vspace{0.5mm}}}
\newcommand{\caseLowTop}[2]{\parbox[b][1.8cm][b]{5cm}{\center #1 \\ \vspace{1mm} #2 \\ \vspace{0.5mm}}}
\newcommand{\caseTallTop}[2]{\parbox[b][2.3cm][b]{5cm}{\center #1 \\ \vspace{1mm} #2 \\ \vspace{0.5mm}}}
\newcommand{\caseLowNext}[2]{\parbox[b][2.2cm][b]{5cm}{\center #1 \\ \vspace{1mm} #2 \\ \vspace{0.5mm}}}
\newcommand{\caseTallNext}[2]{\parbox[b][2.7cm][b]{5cm}{\center #1 \\ \vspace{1mm} #2 \\ \vspace{0.5mm}}}

\begin{document}

\title{Improved mixing bounds for the
Anti-Ferromagnetic Potts
     Model on $\mathbb{Z}^2$
\thanks{This work was partially supported by the EPSRC grant ``Discontinuous Behaviour in the
     Complexity of Randomized Algorithms''.
} }
\author{Leslie Ann Goldberg\\
        Markus Jalsenius\\
        Russell Martin\\
        Mike Paterson\\\\
        Department of Computer Science\\
        University of Warwick\\
        Coventry, CV4 7AL, UK
}

\maketitle
\begin{abstract}
We consider the anti-ferromagnetic Potts model on the the integer
lattice~$\mathbb{Z}^2$.
The model has two parameters, $q$, the number of spins, and
$\lambda=\exp(-\beta)$, where $\beta$ is ``inverse temperature''.
It is known that the model has strong spatial mixing if $q>7$,
or if $q=7$ and $\lambda=0$ or $\lambda>1/8$, or
if $q=6$ and $\lambda=0$ or $\lambda>1/4$.
The $\lambda=0$ case corresponds to the model in which configurations
are proper $q$-colourings of $\mathbb{Z}^2$.
We show that the system has strong spatial mixing for $q\geq 6$ and any~$\lambda$.
This implies that Glauber dynamics is rapidly mixing
(so there is a fully-polynomial randomised approximation scheme
for the partition function)
and
also that there is a unique infinite-volume Gibbs state.
We also show that strong spatial mixing occurs for a larger range
of~$\lambda$ than was previously known for $q=3$, $4$ and $5$.
\end{abstract}

\section{Introduction and statement of results}

\subsection{The anti-ferromagnetic Potts model}

We consider the anti-ferromagnetic Potts model on
the integer lattice~$\mathbb{Z}^2$.
The set of spins is $Q=\{1,\ldots,q\}$. Configurations
are assignments of spins to vertices, and
$\Omega=Q^{\mathbb{Z}^2}$
is the set of all configurations.
A region~$R$ is a (not necessarily connected) subset of vertices,
and $\sigma_R$ denotes the restriction of configuration~$\sigma$ to~$R$.
$\Omega_R=Q^R$ is the set of all such restrictions.
If $R$ is a finite region, then its vertex boundary,
$\partial R$, is the set of vertices that are not in~$R$, but
are adjacent to~$R$.
A boundary configuration on $\partial R$ is a function from
$\partial R$ to the set $\{0\} \cup Q$. The spin ``$0$''
corresponds to a ``free boundary'' which does not influence
the vertices in~$R$.
Let $E(R)$ denote the set of lattice edges that have at least one vertex in~$R$.
Given a region~$R$ and a boundary configuration~$\calB$ on $\partial R$,
the energy of the configuration $\sigma_R\in \Omega_R$ is
given by the Hamiltonian
$$H(\sigma)=\sum_{(i,j)\in E(R)} \beta \delta(\sigma_i ,\sigma_j),$$
where  $\beta\in \mathbb{R}$ is the ``inverse temperature'' and
$$
\delta(s,s')=
\begin{cases}1,&\text{if $s=s'$;}\\
                       0,&\text{otherwise}.
\end{cases}
$$
The {\it partition function\/}
$Z=\sum_{\sigma\in \Omega_R}\exp(-H(\sigma))$.
The {\it finite-volume Gibbs measure\/} $\pi_{\calB}$
is the distribution on $\Omega_R$
in which, for every $\sigma\in \Omega_R$,
$\pi_{\calB}(\sigma)
=\exp(-H(\sigma))/Z$.
Letting $\mon_\sigma(E(R))$ denote
the number of monochromatic edges in $E(R)$ and
taking $\lambda=\exp(-\beta)$,
it is apparent that $\pi_{\calB}(\sigma)$
is proportional to $\lambda^{\mon_\sigma(E(R))}$.

In the zero-temperature case $\beta=\infty$, $\lambda=0$
and $\pi_{\calB}$ is the uniform distribution on ``proper'' colourings,
which are configurations without monochromatic edges.
In this paper we will focus on the situation in which the temperature
is non-zero, so
$\lambda\in(0,1]$.

For any $\Lambda \subseteq R$, $\pi_{\calB,\Lambda}$
denotes the distribution on configurations of $\Omega_\Lambda$ induced by
$\pi_{\calB}$.

\subsection{Strong spatial mixing}

If the parameters $q$ and $\lambda$ are chosen appropriately, then the
anti-ferromagnetic Potts model has {\it strong spatial mixing.}
Informally, this means
that
for any finite region~$R$, if you consider two different
boundary
configurations~$\calB$ and~$\calB'$ on $\partial R$ which differ
at a single vertex~$y$ then
the effect that this difference has on a
subset $\Lambda\subseteq R$ decays exponentially with the distance
from~$\Lambda$ to~$y$.
The formal definition below is taken from~\cite{dsvw} but
adapted to the special case of the anti-ferromagnetic Potts model on $\mathbb{Z}^2$.

\begin{definition}
\label{defssm}
The anti-ferromagnetic Potts model on $\mathbb{Z}^2$ has
strong spatial mixing for parameters~$\lambda$ and~$q$ if
there are constants $\eta$ and $\eta'>0$ such that, for
any non-empty finite region~$R$, any $\Lambda \subseteq R$, any
vertex $y\in
\partial R$, and any pair of boundary configurations $(\calB,\calB')$
of~$\partial R$ which differ only at~$y$,
$$\dtv(\pi_{\calB,\Lambda},\pi_{\calB',\Lambda})
\leq \eta |\Lambda| \exp(-\eta' d(y,\Lambda)),$$ where
$d(y,\Lambda)$ is the lattice distance within~$R$ from the vertex~$y$ to the
region~$\Lambda$ and $\dtv$ denotes total variation distance.

We assume that~$y$ is not a free-boundary vertex in either configuration. That
is, $\calB_y\in Q$ and $\calB'_y\in Q$.
\end{definition}

Strong spatial mixing is an important property because
of two, related, consequences.
First, strong spatial mixing implies that there is a unique
{\it infinite-volume Gibbs measure\/} on configurations in~$\Omega$.
Qualitatively, there is one equilibrium, not many.
Second, strong spatial mixing implies that {\it Glauber dynamics\/}
can be used to efficiently sample configurations from~$\pi_B$
(for any finite region $R$ and  boundary configuration~$\calB$ on~$\partial R$).
We describe both of these consequences below before stating our results.

\subsection{Uniqueness}

A measure~$\mu$ on $\Omega$
is an
infinite-volume {\it Gibbs measure\/}
if, for any finite region~$R$ and any configuration~$\sigma$, the conditional
probability distribution $\mu(\cdot \mid \sigma_{\overline{R}})$
(conditioned on the configuration $\sigma_{\overline{R}}$ on all
vertices other than those in~$R$) is
$\pi_{\sigma_{\partial R}}$.  It is known that there is at least one
infinite-volume Gibbs measure corresponding to any choice of the
parameters~$q$ and~$\lambda$.
An important problem in statistical physics is determining for which
parameters this is unique.
Strong spatial mixing implies that there is a unique infinite-volume Gibbs
measure~\cite{martinelli, dror} with exponentially decaying correlations.

\subsection{Rapid mixing}
\label{sec:rapid}

Suppose that $R$ is a finite region of $\mathbb{Z}^2$ and that $\calB$ is a
boundary configuration on $\partial R$. We will consider the (heat-bath)
Glauber dynamics for sampling from $\pi_\calB$. This is a Markov chain~$\calM$
with state space $\Omega_R$. A transition is made from a configuration
$\sigma\in \Omega_R$ by choosing a vertex~$v$ uniformly at random from~$R$,
``erasing'' the spin at vertex~$v$ and then choosing a new spin for vertex~$v$
from the conditional distribution, given $\sigma_{R-\{v\}}$ and~$\calB$. Here
is a detailed description of the transition.

\smallskip

\noindent {\bf One step of the (heat-bath) Glauber dynamics Markov chain~$\calM$:}
\begin{enumerate}
\item{Choose a vertex $v$ uniformly at random from $R$.}
\item{For $i\in Q$, let $n_i$ denote the number of neighbours
of $v$ which are assigned spin~$i$ (either in~$\sigma$ or in~$\calB$).}
\item{Choose a new spin $c$
according to the distribution
$$\Pr(c = i) = \frac{\lambda^{n_i}}{\sum_{k\in Q} \lambda^{n_k}}$$
for $i\in Q$.}
\item{Obtain the new configuration $\sigma'$ from $\sigma$ by
assigning spin~$c$ to vertex~$v$.}
\end{enumerate}

\smallskip

It is known (for example, see~\cite{dsvw}) that $\calM$ is ergodic, with unique
stationary distribution
$\pi_\calB$\footnote{
It is easy to verify that $\calM$ is ergodic for the positive temperature case
$\lambda \in(0,1]$ considered in this paper. Ergodicity is much more subtle
in the zero-temperature case $\lambda=0$. Here is an example that is not ergodic
with $\lambda=0$ and $q=5$. The region~$R$ consists of two adjacent vertices~$u$
and~$v$. The boundary configuration $\calB$ assigns colours $3$, $4$ and $5$ to the
neighbours of~$u$ and the same colours to the neighbours of~$v$.
Now $\calM$ is not ergodic since it cannot move between the configuration $(u,v)=(1,2)$
and the configuration $(u,v)=(2,1)$.
However, the chain
is ergodic if $q\geq 6$ (the maximum degree plus two) and it is ergodic if $q\geq 3$
if the boundary configuration is chosen appropriately (for example, the
free boundary case). See, for example, the ergodicity proofs in~\cite{gmp, lrs}.
}.
It is also known that if the Potts model has strong
spatial mixing (which is true for appropriate choices of~$q$ and~$\lambda$, as
we will see below) then~$\calM$ is rapidly mixing.

Before describing what is known about rapid mixing, we recall
the definitions. Let $P$ denote the
transition matrix of~$\calM$, and let $P^t(\sigma,\sigma')$
be the $t$-step probability of moving from $\sigma$ to $\sigma'$.
For $\delta > 0$, the mixing time is defined as
$\tau_{\calM}(\delta) = \min \{ t_0 : d_{tv}(P^t,\pi_{\calB})
     \leq \delta \textrm{ for all } t \geq t_0\}$.
$\calM$ is said to be {\em rapidly mixing} if
$\tau_{\calM}(\delta)$ is
at most a polynomial in~$n$ and  $\log(1/\delta)$, where $n$ is the number
of vertices in $R$.

It is well-known that strong spatial mixing implies rapid mixing in our
setting. The difficulty of the proof depends upon the precise bound on
$\tau_\calM(\delta)$ that is obtained. Dyer, Sinclair, Vigoda and
Weitz~\cite[Theorem 2.5]{dsvw} give a nice simple combinatorial proof that
strong spatial mixing implies that a certain ``heat-bath block dynamics'' mixes
in $O(n \log(n/\delta))$ time. Markov-chain comparison can now be applied in a
standard way to show that Glauber dynamics mixes in $O(n(n+\log(1/\delta)))$
time (see, for example, Section~7 of~\cite{gmp} or (for a slightly larger
bound)~\cite{six}). In fact, it is known that strong spatial mixing implies
$O(n \log(n/\delta))$ mixing of Glauber dynamics, giving a small improvement on
the mixing-time bound. As explained in~\cite{dsvw}, this can be proved using
techniques from functional analysis~\cite{cesi, martinelli, mo, sz}. The idea
is to bound the log-Sobolev constant of the block dynamics, and translate this
bound into a bound on the log-Sobolev constant of Glauber dynamics.

\subsection{Approximating the partition function}
\label{sec:fpras}

We have seen in Section~\ref{sec:rapid} that when the Potts model has
strong spatial mixing, the Markov chain~$\calM$, which corresponds to heat-bath
Glauber dynamics, is rapidly mixing. Thus, there is an efficient algorithm
for sampling from the Gibbs distribution $\pi_{\calB}$.

Before stating our results in Section~\ref{sec:context},
we mention one consequence of rapid mixing.
A \emph{randomised approximation scheme\/} is an algorithm for
approximately computing the value of a function~$f$. The
approximation scheme has a parameter~$\varepsilon>0$ which specifies
the error tolerance. For concreteness, suppose that~$f$ is a
function from~$\Sigma^*$ to~$\mathbb{R}$. For example,
for fixed values of~$q$ and~$\lambda$,
$f$
might map an encoding of
a region~$R$ and a boundary configuration $\calB$ to the value of the partition
function~$Z$ corresponding to~$R$ and~$\calB$.
A \emph{randomised approximation scheme\/} for~$f$ is a
randomised algorithm that takes as input an instance $ x\in
\Sigma^*$  (e.g., $R$ and $\calB$) and an error
tolerance $\varepsilon >0$, and outputs a number $z\in \mathbb{Q}$
(a random variable of the ``coin tosses'' made by the algorithm)
such that, for every instance~$x$,
\begin{equation}
\label{eq:3:FPRASerrorprob}
\Pr \left[\frac{f(x)}{
1+\epsilon}\leq z \leq (1+\epsilon) f(x)\right]\geq \frac{3}{4}\, .
\end{equation}
The randomised approximation scheme is said to be a
\emph{fully polynomial randomised approximation scheme},
or \emph{FPRAS},
if it runs in time bounded by a polynomial
in $ |x| $ and $ \epsilon^{-1} $.
Note that the quantity $3/4$ in
Equation~(\ref{eq:3:FPRASerrorprob})
could be changed to any value in the open
interval $(\frac12,1)$ without changing the set of problems
that have randomised approximation schemes.

Using ideas of Jerrum, Valiant and Vazirani~\cite{jvv},
an efficient sampling algorithm for~$\pi_{\calB}$
can be turned into an FPRAS for the partition function.
A straightforward proof is based on Dyer and Greenhill's
extension~\cite{dg} of~\cite{jvv}.

In summary, if $q$ and $\lambda$ are chosen so that
the Potts model has strong spatial mixing then
$\calM$ is rapid mixing. This, in turn, gives an FPRAS for the partition
function.

\subsection{Context and statement of results}
\label{sec:context}

For $q=2$ (see~\cite{martinelli}) it is known that there is a critical point
$\lambda_c$ such that uniqueness (and strong spatial mixing) hold for
$\lambda>\lambda_c$ but there are two Gibbs measures for $\lambda<\lambda_c$
(in one of these Gibbs measures, spin~$1$ is favoured at ``even-parity''
vertices, and in the other, spin~$2$ is favoured). The value of $\lambda_c$
(see~\cite{ss}) is $\lambda_c=\sqrt{2}-1\sim 0.41$.

Thus, we investigate the case $q>2$. It is believed~\cite{ss} that there is
strong spatial mixing for every $\lambda\in(0,1]$ for $q=3$ and for every
$\lambda\in[0,1]$ for $q>3$.
The point $q=3$, $\lambda=0$ is excluded because, on physical grounds, this is believed
to be a critical point. It is believed that at this point there is a unique
infinite-volume Gibbs measure but that the correlations only decay
algebraically (e.g., polynomially).
Salas and Sokal used {\it Dobrushin uniqueness\/}
to show that that strong spatial mixing occurs for every $\lambda\in[0,1]$ for
$q>8$. As Jerrum points out~\cite[Section 5]{jerrum}, Salas and Sokal's
calculation applies whenever $q>8(1-\lambda)$, so it also applies to
positive~$\lambda$ for smaller~$q$. The result applies to a more general
context than the one studied in this paper --- it applies to the
anti-ferromagnetic Potts model on any infinite graph. The generalised condition
is $q>2\Delta(1-\lambda)$, where $\Delta$ is the maximum degree.
Jerrum~\cite{jerrum} considered the $\lambda=0$ case and showed rapid mixing
(in fact, $O(n \log(n/\delta))$ mixing) for Glauber dynamics when $q>2\Delta$
(in fact, he considered a slightly different version of Glauber dynamics, but
the difference is not important here). Jerrum's result implies Salas and
Sokal's for $\lambda=0$ since $O(n \log(n/\delta))$ mixing of Glauber dynamics
implies strong spatial mixing~\cite[Theorem 2.3]{dsvw}.

The results that we have just discussed give strong spatial mixing for $\lambda=0$
and $q>8$.
In fact, better results are known for $\lambda=0$.
Salas and Sokal~\cite{ss} used {\it decimation\/} to prove strong spatial mixing
for $q\geq 7$. This is a machine-assisted proof.
The $q=7$ case is also implied by the work of Bubley, Dyer, Greenhill and Jerrum~\cite{bdgj}.
They gave a machine-assisted proof of
$O(n \log(n/\delta))$ mixing for a block dynamics on $4$-regular triangle-free
graphs. As we mentioned above, this implies
$O(n \log(n/\delta))$ mixing for Glauber dynamics, which, in turn, implies
strong spatial mixing.
A proof without machine assistance
of strong spatial mixing for $q\geq 7$
is given by Goldberg, Martin and Paterson~\cite[Theorem 5]{gmp}.
Once again, the result applies more generally --- in this case to
triangle-free graphs with maximum degree at most~$\Delta\geq 3$
where $q>1.76\Delta-0.47$.

Achlioptas et al.~\cite{six} gave a machine-assisted proof of strong spatial
mixing for $\lambda=0$ and $q=6$. Their method was to prove $O(n
\log(n/\delta))$ mixing for a block dynamics, which implies spatial mixing as
discussed above.

It is known that Glauber dynamics is rapidly mixing
on rectangular regions when $q=3$ and $\lambda=0$.
This is proved in the fixed-boundary case by Luby, Randall, and Sinclair~\cite{lrs}
and in the free-boundary case by Goldberg, Martin and Paterson~\cite{gmpthree}.
The (polynomial) mixing-time bounds are not
$O(n \log(n/\delta))$. Indeed, as mentioned above, it is not believed
that strong spatial mixing holds for $\lambda=0$ and $q=3$.

The following proposition summarises the results that we have just discussed.
\begin{proposition}\label{prop:known}
Consider the anti-ferromagnetic Potts model on $\mathbb{Z}^2$
with parameters~$q$ and $\lambda\leq 1$.
There is strong spatial mixing in the following cases.
\begin{enumerate}
\renewcommand{\theenumi}{(\roman{enumi})}
\renewcommand\labelenumi{\theenumi}
\item{$q\geq8$ and $\lambda\geq 0$,}
\item{$q= 7$ and $\lambda=0$ or $\lambda>1/8=0.125$,}
\item{$q = 6$ and $\lambda=0$ or $\lambda> 2/8=0.25$,}
\item{$q = 5$ and $\lambda > 3/8=0.375$,}
\item{$q = 4$ and $\lambda > 4/8=0.5$, and}
\item{$q = 3$ and $\lambda > 5/8=0.625$.}
\end{enumerate}
Thus, in these cases, Glauber dynamics is rapidly mixing
and there is a unique infinite-volume Gibbs measure.
\end{proposition}

The purpose of this work is to improve the
results in Proposition~\ref{prop:known}.
Our main objective was to extend the $q=6$ and $q=7$ results
for $\lambda=0$ to all temperatures.
We state our results as two theorems to separate the results
that are proved without machine assistance (Theorem~\ref{thm:no-computer})
from those that are proved with machine assistance.
Theorem~\ref{thm:computer} subsumes Theorem~\ref{thm:no-computer}.

\begin{theorem}\label{thm:no-computer}
Consider the anti-ferromagnetic Potts model on $\mathbb{Z}^2$
with parameters~$q$ and $\lambda\leq 1$.
There is strong spatial mixing in the following cases.
\begin{enumerate}
\renewcommand{\theenumi}{(\roman{enumi})}
\renewcommand\labelenumi{\theenumi}
\item{$q\geq 7$ and $\lambda\geq 0$,}
\item{$q = 6$ and $\lambda=0$ or $\lambda > 1/7\approx 0.1429$,}
\item{$q = 5$ and $\lambda > 2/7 \approx 0.2857$,}
\item{$q = 4$ and $\lambda > \frac{1}{2}(\sqrt{33} - 5)\approx 0.3723$, and}
\item{$q = 3$ and $\lambda > \lambda_0$, where $\lambda_0\approx 0.4735$ is
     the real solution of $\lambda^3+4\lambda-2=0$.}
\end{enumerate}
Thus, in these cases, Glauber dynamics is rapidly mixing
and there is a unique infinite-volume Gibbs measure.
\end{theorem}

\begin{theorem}\label{thm:computer}
Consider the anti-ferromagnetic Potts model on $\mathbb{Z}^2$
with parameters~$q$ and $\lambda\leq 1$.
There is strong spatial mixing in the following cases.
\begin{enumerate}
\renewcommand{\theenumi}{(\roman{enumi})}
\renewcommand\labelenumi{\theenumi}
\item{$q \geq 6$ and $\lambda \geq 0$,}\label{thm:part-6comp}
\item{$q = 5$ and $\lambda \geq 0.127$,}
\item{$q = 4$ and $\lambda \geq 0.262$, and}
\item{$q = 3$ and $\lambda \geq 0.393$.}
\end{enumerate}
\end{theorem}

The bounds for $q=5$, $q=4$ and $q=3$ can be improved further
by more extensive machine calculation. These results
will appear in the PhD thesis of one of the authors~\cite{jalsenius}.

\subsection{The anti-ferromagnetic Potts model on general graphs}

In this paper we consider the anti-ferromagnetic Potts model
on the integer lattice~$\mathbb{Z}^2$.
One reason for restricting attention to~$\mathbb{Z}^2$ is
that it is a natural lattice, of interest in statistical physics~\cite{martinelli}.
Another reason is that the model is known {\it not\/} to have good
mixing properties on a general graph.
As Welsh observes~\cite[3.7.12]{welsh},
the partition function of the Potts model is a specialisation
of the {\it Tutte Polynomial\/} along the hyperbola
$H_q=\{(x,y): (x-1)(y-1)=q\}$.
The {\it anti-ferromagnetic\/} Potts model
(for real temperatures)
corresponds to the additional constraint $0\leq \lambda\leq 1$,
which corresponds to a portion of the hyperbola
in which $x-1$ and $y-1$ are negative.
There is no FPRAS for the Tutte polynomial along this hyperbola unless
NP=RP~\cite[8.7.2]{welsh}.

Jerrum and Sinclair~\cite{ising} considered the anti-ferromagnetic
Ising model, which corresponds to the Potts model with $q=2$.
They used a reduction from {\sc MaxCut} (the problem of counting cut-sets
of a specified size
in a graph) to show that there is no FPRAS
for the partition function unless NP=RP.
Their proof applies for a particular value of~$\lambda$,
but the stretching and thickening technique of Jaeger, Vertigan and
Welsh~\cite{jvw} can be used to show that there is no FPRAS
for any fixed~$\lambda$ (see~\cite{twospins}).
Welsh has shown that the same result holds for any $q\geq 3$~\cite[8.7.2]{welsh}.
Thus, unless NP=RP, the anti-ferromagnetic model does not
exhibit strong spatial mixing on a general graph.
In this paper, we do not consider a general graph. Instead we consider
the integer lattice~$\mathbb{Z}^2$.

\section{Recursive coupling}
\subsection{The recursive coupling tree}

The essence of proving strong spatial mixing is showing that, if you take an
arbitrary region~$R$ and boundary configurations~$\calB$ and~$\calB'$ on
$\partial R$ that disagree on a single boundary vertex~$y$, then there is a
coupling of~$\pi_\calB$ and~$\pi_{\calB'}$ in which the probability of
disagreement at a vertex decays exponentially with its distance from~$y$. We
will construct such a coupling using the recursive method of Goldberg, Martin
and Paterson~\cite{gmp}. We start by describing the method.

Let $R$ be a non-empty finite region. As in~\cite{gmp}, we will find it
convenient to work with the edge-boundary of~$R$ rather than with the boundary
$\partial R$ of vertices surrounding~$R$. Here is the notation that we will
use. The {\it boundary\/} of the region~$R$ is the collection of edges that
have exactly one endpoint in~$R$. A {\it boundary configuration}~$B$ is a
function from the set of edges in the boundary to the set $\{0\}\cup Q$. Given
a configuration $\sigma\in \Omega_R$, the quantity $\mon_\sigma(E(R))$ is the
number of monochromatic edges in~$E(R)$, where a boundary edge is said to be
``monochromatic'' if its spin is the same as the spin that is assigned
by~$\sigma$ to its endpoint. $\pi_B$ is the Gibbs distribution in which the
probability of~$\sigma$ is proportional to $\lambda^{\mon_\sigma(E(R))}$. We
will be interested in studying how much $\pi_B$ varies when we change the spin
of a single edge of~$B$. This small change to the boundary is formalised by the
following notation.

\begin{definition}
\label{def:bp}
A {\it boundary pair\/}\footnote{In the paper~\cite{gmp}, this was referred to
as a ``relevant boundary pair''. The reason for the terminology
is that paper~\cite{gmp} also used the notion of a boundary pair in which the
final condition above (the one about perpendicular boundary edges) is dropped.
Note that this condition depends upon the geometry of the lattice.
In this paper we always work on the lattice $\mathbb{Z}^2$ and we always include
all conditions listed above so
we drop the word ``relevant'' to simplify terminology.} $X$ consists of
\begin{itemize}
\item a non-empty finite region~$R_X$,
\item a distinguished boundary edge $s_X=(w_X,f_X)$ with $f_X\in R_X$, and
\item a pair $(B_X,B'_X)$ of boundary configurations
which differ only on the edge~$s_X$.
\end{itemize}
We require
\begin{itemize}
\item $B_X(s_X)\in Q$, and
\item $B'_X(s_X)\in Q$, and
\item  any two perpendicular
boundary edges that share a vertex $f\in\partial R_X$ have the same spin in at
least one of the two configurations $B_X$ and $B'_X$.
\end{itemize}
\end{definition}

A {\it coupling\/} $\Psi$
of~$\pi_{B_X}$ and~$\pi_{B'_X}$
is a distribution on $\Omega_{R_X}\times \Omega_{R_X}$
which has marginal distributions~$\pi_{B_X}$ and~$\pi_{B'_X}$.
For such a coupling~$\Psi$, we define
$1_{\Psi,f}$ to be the indicator random variable for the event that,
when a pair of configurations is drawn from~$\Psi$,
the spin of~$f$ differs in these two configurations.
For any boundary pair $X$ we define $\Psi_X$ to be some coupling of~$\pi_{B_X}$
and~$\pi_{B'_X}$ minimizing $\expected[1_{\Psi,f_X}]$. For every pair of
spins~$c$ and~$c'$, let $p_X(c,c')$ be the probability that, when a
pair of configurations $(C,C')$ is drawn from~$\Psi_X$,
$f_X$ has spin~$c$ in~$C$ and spin~$c'$ in~$C'$.

We define a labelled tree~$T_X$
associated with each boundary pair~$X$.
We will use the tree to get an upper bound on
the expected number of disagreements at any distance from~$w_X$ in a coupling of
$\pi_{B_X}$ and $\pi_{B'_X}$.

The tree $T_X$ is constructed as follows. Start with a vertex~$r$ which will be
the root of~$T_X$. For every pair of spins $c\in Q$ and~$c'\in Q$, $c \neq c'$,
add an edge labelled $(p_X(c,c'),f_X)$ from~$r$ to a new node $r_{c,c'}$. If
$f_X$ has no neighbours in~$R_X$ then $r_{c,c'}$ is a leaf. Otherwise, for some
$k\in\{1,2,3\}$, let $e_1,\ldots,e_k$ be the edges from~$f_X$ to nodes
in~$R_X$.
If $k=3$ order these edges so that $e_1$ and $e_3$ are not perpendicular.
For each $i\in\{1,\ldots,k\}$, let $X_i(c,c')$ be the
boundary pair consisting of
\begin{itemize}
\item the region $R_X - f_X$;
\item the distinguished edge $e_i$;
\item the boundary configuration~$B$ of $R_X - f_X$ that
\begin{itemize}
\item agrees with $B_X$ on common edges,
\item assigns spin~$c'$ to  $e_1,\ldots,e_{i-1}$, and
\item assigns spin~$c$ to $e_i,\ldots,e_k$; and
\end{itemize}
\item the boundary configuration~$B'$ that agrees with $B$ except that it
assigns spin~$c'$ to~$e_i$.
\end{itemize}
Recursively construct $T_{X_i(c,c')}$, the tree corresponding to boundary
pair~$X_i(c,c')$. Add an edge with label $(1,\cdot)$ from~$r_{c,c'}$ to
the root of~$T_{X_i(c,c')}$.
That completes the construction of~$T_X$.

We say that an edge~$e$ of~$T_X$ is {\it degenerate\/} if the
second component of its label is ``$\cdot$''.
For edges $e$ and $e'$ of $T_X$, we write $e\rightarrow e'$
to denote the fact that $e$ is an ancestor of~$e'$.
That is, either $e=e'$, or $e$ is a proper ancestor of~$e'$.
Define the {\it level\/} of edge~$e$
to be the number of non-degenerate edges on the path from
the root down to, and including, $e$.
Suppose that
$e$ is an edge of~$T_X$ with label~$(p,f)$.
We say that the {\it weight\/} $w(e)$ of edge~$e$ is~$p$.
Also the {\it name\/} $n(e)$ of edge~$e$ is~$f$.
The {\it likelihood\/} $\ell(e)$ of~$e$
is $\prod_{e':e'\rightarrow e} w(e)$.
The {\it cost\/} $\gamma(f,T_X)$
of a vertex~$f$ in~$T_X$
is $\sum_{e:n(e)=f}\ell(e)$.
For any $d\geq 1$, let $E_d(X)$ denote the set of level-$d$ edges in~$T_X$.
Let $\Gamma_d(X) = \sum_{e \in E_d(X)} \ell(e)$.
We use the following lemma, from~\cite{gmp}.
\begin{lemma}\cite{gmp}
Consider the anti-ferromagnetic Potts model on $\mathbb{Z}^2$ with
parameters~$q$ and~$\lambda$.
Suppose there is an $\epsilon>0$ such that, for every boundary pair~$X$
and every $d\geq 1$, $\Gamma_d(X)\leq{(1-\epsilon)}^{d}$.
Then the system has strong spatial mixing.
\label{lem:tree}
\end{lemma}

\begin{proof}
The relevance of~$T_X$ for providing an upper bound on the quality of the
coupling is established in Lemma~12 of~\cite{gmp}, which shows that there is a
coupling~$\Psi$ of $\pi_{B_X}$ and $\pi_{B'_X}$ such that, for all $f\in R_X$,
$\expected[1_{\Psi,f}] \leq\gamma(f,T_X)$ which is at most $\sum_{d\geq
d(f,s_x)} \Gamma_d(X)$, where $d(f,s_X)$ is the lattice distance from~$f$
to~$s_X$. (Thus, $d(f_X,s_x)=1$ and if $f\in R_X$ is adjacent to~$f_X$ then
$d(f,s_X)=2$ and so on.)
Following the proof of~\cite[Lemma 33]{gmp}, we find
that
$$\expected[1_{\Psi,f}] \leq
\frac{1}{\epsilon}
{(1-\epsilon)}^{d(f,s_X)}$$
and
$$\sum_{f\in R_X}\expected[1_{\Psi,f}] \leq\frac{1-\epsilon}{\epsilon}.$$
Following the proof of \cite[Lemma 34]{gmp}, we obtain  similar conclusions, assuming that
we start with a pair of boundary configurations on the boundary $\partial R$ of
{\it vertices\/} surrounding~$R$, such that the pair differs only on a
particular vertex~$v_X$. In particular, there is a coupling~$\Psi$ such that
$$\expected[1_{\Psi,f}] \leq
\frac{6}{\epsilon(1-\epsilon)} {(1-\epsilon)}^{d(f,v_X)}$$ and
$$\sum_{f\in R_X}\expected[1_{\Psi,f}] \leq\frac{6}\epsilon.$$
This implies strong spatial mixing~\cite[Corollary 21]{gmp}.
\end{proof}

\subsection{Bounding the cost of level-$d$ edges in the recursive coupling tree}

A key ingredient from the construction of~$T_X$ which affects $\gamma(f,T_X)$
is the quantity $\expected[1_{\Psi_X,f_X}]$, which we denote $\nu(X)$. Thus,
$\nu(X)=\min_\Psi \expected[1_{\Psi,f_X}]$, where the minimum is over all
couplings~$\Psi$ of~$\pi_{B_X}$ and $\pi_{B'_X}$.

In order to get good upper bounds on $\nu(X)$, Goldberg, Martin and
Paterson~\cite{gmp} observed that $\nu(X)$ can be upper-bounded in terms of
corresponding values for boundary pairs with smaller regions. They used the
following lemma.
\begin{lemma}\cite{gmp}
\label{lem:oldmiss} Suppose $\lambda=0$. Suppose that $X$ is a boundary pair.
Let $R'$ be any subset of $R_X$ which includes~$f_X$. Let $\chi$ be the set of
boundary pairs~$X'=(R_{X'},s_{X'},B_{X'},B'_{X'})$ such that $R_{X'}=R'$,
$s_{X'}=s_X$, $B_{X'}$ agrees with $B_X$ on common edges, and $B'_{X'}$ agrees
with $B'_X$ on common edges. Then $\nu(X) \leq \max_{X'\in \chi} \nu(X')$.
\end{lemma}

\begin{figure}[ht]
\begin{center}
\setlength{\unitlength}{0.2cm}
\small 
\begin{picture}(23, 15)
\qbezier(4,3)(0,4)(1,8)
\qbezier(1,8)(2,12)(6,14)
\qbezier(6,14)(10,16)(11,14)
\qbezier(11,14)(14,8)(17,12)
\qbezier(17,12)(20,16)(22,12)
\qbezier(22,12)(23,10)(22,6.5)
\qbezier(22,6.5)(21,3)(17,2)
\qbezier(17,2)(12,1)(4,3)
\put(13,0){\line(0,1){4}}
\put(13,0){\circle*{0.5}}
\put(13,4){\circle*{0.5}}
\put(7,11){\makebox{$R_X$}}
\put(10.2,4){\makebox{$f_X$}}
\put(9.8,0){\makebox{$w_X$}}
\put(13.5,2.6){\makebox{$s_X$}}
\end{picture}
\hspace{1cm}
\begin{picture}(23, 15)
\qbezier(4,3)(0,4)(1,8)
\qbezier(1,8)(2,12)(6,14)
\qbezier(6,14)(10,16)(11,14)
\qbezier(11,14)(14,8)(17,12)
\qbezier(17,12)(20,16)(22,12)
\qbezier(22,12)(23,10)(22,6.5)
\qbezier(22,6.5)(21,3)(17,2)
\qbezier(17,2)(12,1)(4,3)
\qbezier(4,3)(4,7)(6.5,8)
\qbezier(6.5,8)(9,9)(13,7)
\qbezier(13,7)(17,5)(17,2)
\put(13,0){\line(0,1){4}}
\put(13,0){\circle*{0.5}}
\put(13,4){\circle*{0.5}}
\put(7,11){\makebox{$R_X$}}
\put(7,5){\makebox{$R'$}}
\put(10.2,4){\makebox{$f_X$}}
\put(9.8,0){\makebox{$w_X$}}
\put(13.5,2.6){\makebox{$s_X$}}
\end{picture}
\caption{The application of Lemma~\ref{lem:oldmiss}.}\label{boundmufig}
\end{center}
\end{figure}
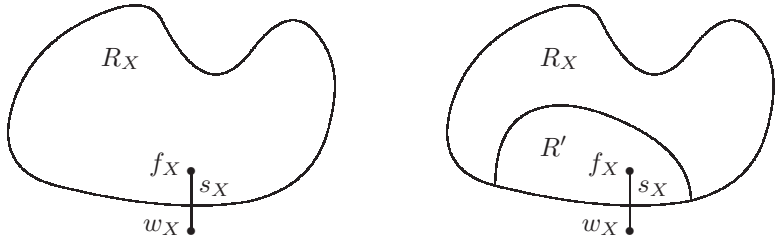

Figure~\ref{boundmufig} is an illustration of how Lemma~\ref{lem:oldmiss} is
used to find an upper bound on~$\nu(X)$. The basic idea is to pick a small
subregion $R'$ that contains the vertex $f_X$. Compute the maximum value of
$\nu$ for that subregion, where we maximise over boundary configurations of
$R'$ that agree with the boundary configurations of~$R_X$ on the common overlap
of these boundaries. This maximum value is an upper bound for $\nu(X)$.

An interesting feature of the positive-temperature Potts model is that this
approach does not work. In particular, Lemma~\ref{lem:oldmiss} does not apply
to positive~$\lambda$. For example, suppose $q=2$ and $\lambda=1/2$. Consider a
region~$R_X$ containing~$f_X$ and one of its neighbours, $y$, as illustrated
below. (In this diagram and all subsequent
diagrams  we will denote vertices as squares so that we have
space to label them.) In the diagram, $s_X$ is the edge between~$f_X$ and its
lower neighbour (which is not pictured). The edge $s_X$ is assigned spins~$1$
and~$2$ by the two boundary configurations~$B_X$ and~$B'_X$. The rest of the
boundary configurations are as shown (assigning spins $1$, $1$, $1$, $2$ and
$2$ clockwise around the picture). A calculation shows that $\nu(X)=30/91$.
However, if $R'$ is chosen to be the region containing~$f_X$ only then the
corresponding boundary pairs~$X'$ (depicted to the right) both have
$\nu(X')=30/100<30/91$.
\bigskip
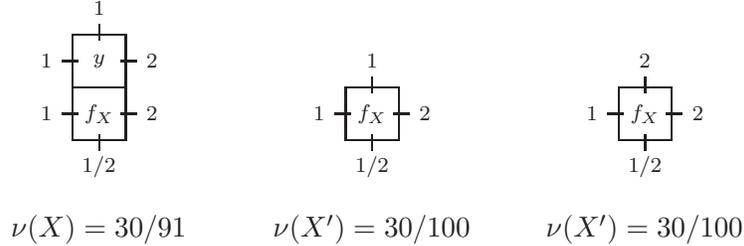
\begin{center}
\setlength{\unitlength}{20pt} \caseConvex{\ConvexBig}{$\nu(X) = 30/91$}
\caseConvex{\ConvexSmallA}{$\nu(X') =30/100$}
\caseConvex{\ConvexSmallB}{$\nu(X') = 30/100$}
\end{center}
\bigskip

Our approach is to find an upper bound, $\mu(X)$, for $\nu(X)$ such
that $\mu(X)$ can be upper-bounded using smaller regions along the lines of
Lemma~\ref{lem:oldmiss}. Let $X$ be a boundary pair. Recall that $E(R_X)$ is
the set of lattice edges with at least one endpoint in~$R_X$. For any subset
$E\subseteq E(R_X) - \{ s_X \}$ and any configuration~$\sigma\in\Omega_{R_X}$,
let $\mon_{\sigma}(E)$ denote the number of monochromatic edges in $E$, where a
boundary edge is considered to be monochromatic if its spin in $B_X$ is the
same as the spin assigned by~$\sigma$ to its endpoint.  For $i\in Q$, let
$\Omega_i$ be the set of configurations in $\Omega_{R_X}$ that assign spin~$i$
to vertex~$f_X$. Let~$c_i$ be the total weight of these configurations,
ignoring edge~$s_X$.
$$ c_i  = \sum_{\sigma\in \Omega_i}
\lambda^{\mon_{\sigma}(E(R_X)-\{s_X\})}.$$ Let~$C$ contain the two spins
assigned to~$s_X$ by the boundary configurations. That is, $C=\{B_X(s_X),B'_X(s_X)\}$
and let $c = \sum_{i\not\in C} c_i$. We now define
$$
\mu(X) = \max_{i\in C} \frac{(1-\lambda)c_i}{(1+\lambda)c_i+c}.
$$
The following lemma enables us to use $\mu(X)$ to find upper bounds for
$\nu(X)$.
The intuition behind the lemma is best understood from Equations~(\ref{newrefeq})
and~(\ref{newnewrefeq}).
Informally, (\ref{newrefeq}) says that the disagreement probability $\nu(X)$ is at most
the difference between the probability of seeing a certain colour in
one distribution (with one boundary configuration) and the probability of
seeing the same colour in the other distribution. A little manipulation gives
Equation~(\ref{newnewrefeq}), which shows that this quantity is at most~$\mu(X)$.
The remainder of the argument shows that $\mu(X)$ can be upper bounded using smaller
regions.\footnote{To see that it is plausible that $\mu(X)$ can be upper bounded using smaller regions,
consider the boundary configuration $B$ which is the same as the 
boundary configurations in~$X$ except that $B(s_X)=0$ so $s_X$ is a free boundary edge.
Note that in the expression
$$\frac{(1-\lambda)c_i}{(1+\lambda)c_i+c} =
\frac{1-\lambda}{1+\lambda+\frac{c}{c_i}},
$$ 
from the definition of $\mu(X)$, $c/c_i$ is the ratio of
$\Pr_{\pi_B}(f_X\not\in C)$ to $\Pr_{\pi_B}(f_X=i)$. By convexity, this ratio can be bounded by 
considering smaller regions (see the proof for details). Of course, the convexity argument
allows some flexibility in the exact definition of $\mu(X)$ and the best thing is
to define $\mu(X)$ so that it is as small as possible, subject to the constraint $\nu(X)\leq\mu(X)$.
. }

\begin{lemma}
\label{lemma:miss} Suppose that $X$ is a boundary pair. Let $R'$ be any subset
of $R_X$ which includes $f_X$. Let $\chi$ be the set of boundary pairs
$X'=(R_{X'}, s_{X'}, B_{X'}, B'_{X'})$ such that $R_{X'}=R', s_{X'}=s_X,
B_{X'}$ agrees with $B_X$ on common edges, and $B'_{X'}$ agrees with $B'_X$ on
common edges. Then $\nu(X) \leq \max_{X' \in \chi} \mu(X')$.

\end{lemma}
\begin{proof}

Without loss of generality (to simplify notation) suppose $B_X(s_X)=1$,
$B'_X(s_X)=2$, and $c_1\geq c_2$. We will show
\begin{enumerate}
\renewcommand{\theenumi}{(\roman{enumi})}
\renewcommand\labelenumi{\theenumi}
\item{\quad $\nu(X) \leq \mu(X)$ and}\label{one}
\item{\quad $\mu(X) \leq \text{max}_{X' \in \chi} \mu(X')$.}\label{two}
\end{enumerate}

First, we show~\ref{one}. Note that
\begin{align*}
\text{Pr}_{\pi_{B_X}}(f_X = i) &=
    \begin{cases}
        \frac{\lambda c_1}{\lambda c_1 + c_2 + c}, & i = 1; \\
        \frac{c_i}{\lambda c_1 + c_2 + c}, & 2 \leq i \leq q,
    \end{cases} \\
\text{Pr}_{\pi_{B_X'}}(f_X = i) &=
    \begin{cases}
        \frac{\lambda c_2}{c_1 + \lambda c_2 + c}, & i = 2; \\
        \frac{c_i}{c_1 + \lambda c_2 + c}, & i = 1, 3 \leq i \leq q.
    \end{cases}
\end{align*}
Since $c_1\geq c_2$ and $\lambda\leq 1$, the denominator in the expression for
$\text{Pr}_{\pi_{B_X'}}(f_X = i)$ exceeds the denominator in
$\text{Pr}_{\pi_{B_X}}(f_X = i)$ so we can couple~$\pi_{B_X}$ and~$\pi_{B'_X}$
in such a way that disagreement at~$f_X$ occurs only when the sample
from~$\pi_{B'_X}$ assigns spin~$1$ to~$f_X$. Thus,
\begin{eqnarray}
\label{newrefeq}
  \nu(X) &\leq& \text{Pr}_{\pi_{B'_X}}(f_X=1)-\text{Pr}_{\pi_{B_X}}(f_X=1) \\
  \nonumber
   &=& \frac{c_1}{c_1 + \lambda c_2 + c} \ - \ \frac{\lambda c_1}{\lambda c_1 + c_2 + c} \\
   \nonumber
   &=& \frac{c_1(1-\lambda)(c_2+\lambda c_2+c)}{(c_2+\lambda c_1+c)(c_1+\lambda c_2+c)} \\
   \label{newnewrefeq}
&\leq& \frac{(1-\lambda)c_1}{(1+\lambda)c_1+c} \leq \mu(X).
\end{eqnarray}

For~\ref{two}, let $W = R_X - R'$. For $i\in Q$ and $\rho\in \Omega_W$ let
$\Omega_{i,\rho}$ be the set of configurations $\sigma\in \Omega_{R_X}$ with
$\sigma_{f_X}=i$ and $\sigma_W=\rho$. Let
$$ c_{i,\rho}  = \sum_{\sigma\in \Omega_{i,\rho}}
\lambda^{\mon_{\sigma}(E(R_X)-\{s_X\})},$$
let $\hat{c}_\rho=\max(c_{1,\rho},c_{2,\rho})$
and let $c_\rho = \sum_{i=3}^q
c_{i,\rho}$. Then
\begin{eqnarray*}
\mu(X) &=& \text{max}\left(\frac{(1-\lambda)c_1}{(1+\lambda)c_1+c},\;
                             \frac{(1-\lambda)c_2}{(1+\lambda)c_2+c}\right)
= \frac{1-\lambda}{1+\lambda+\frac{c}{c_1}}
= \frac{1-\lambda}
{1+\lambda+
      \frac {\sum_{\rho \in \Omega_W} c_\rho\strut}
            {\strut \sum_{\rho \in \Omega_W} c_{1,\rho}}}\\
&\leq& \frac{1-\lambda}
{1+\lambda+
       \frac{\sum_{\rho \in \Omega_W} c_\rho\strut}
       {\strut \sum_{\rho \in \Omega_W} \hat{c}_\rho
                        }}
= \frac{(1-\lambda)\sum_{\rho \in \Omega_W}
\hat{c}_\rho \strut
               }{\strut (1+\lambda)\sum_{\rho \in \Omega_W}
\hat{c}_\rho
                +\sum_{\rho \in \Omega_W}c_\rho}
= \frac{\sum_{\rho \in \Omega_W}(1-\lambda)
\hat{c}_\rho\strut
               }{\strut \sum_{\rho \in \Omega_W}((1+\lambda)
\hat{c}_\rho
               +c_\rho)} \\
&\leq& \max_{\rho \in \Omega_W} \frac{(1-\lambda)
\hat{c}_\rho
}{(1+\lambda)
\hat{c}_\rho
+c_\rho}
= \max_{\rho \in
\Omega_W}\left(\text{max}\left(\frac{(1-\lambda)c_{1,\rho}}{(1+\lambda)c_{1,\rho}+c_\rho},\;
       \frac{(1-\lambda)c_{2,\rho}}{(1+\lambda)c_{2,\rho}+c_\rho}\right)\right)\\
&=& \max_{\rho \in \Omega_W} \mu(X'),
\end{eqnarray*}
where $X'$ is the boundary pair in $\chi$ in which
$B_{X'}$ and $B'_{X'}$ are induced by $\rho$.
Note that $X'$ is a boundary pair --- in particular, it satisfies the
condition about perpendicular edges.
The last step follows from the observation that
$c_{1,\rho}$, $c_{2,\rho}$ and $c_\rho$ all contain the factor
$\lambda^{\text{mon}_\sigma(E(R_X)-E(R_{X'}))}$, which is constant for a fixed~$\rho$,
and can be cancelled out to obtain $\mu(X')$.
\end{proof}

\section{Proof of Theorem~\ref{thm:no-computer}}
\label{sec:big}

We start with a lemma, which we will use to obtain upper bounds on~$\mu(X)$.
The intuition behind the lemma is that if $R_X$ is the region consisting of a single
node~$f_X$ then $\mu(X)$ is maximised by avoiding the colours of $s_X$ in the
rest of the boundary and otherwise spreading colours evenly over the boundary.

\begin{lemma}
Suppose that $X$ is a boundary pair in which $R_X$ consists of a node
$f_X$ only.

Let $v=3\mod (q-2)$ and $u=\lfloor 3/(q-2)\rfloor$. (So
$u(q-2)+v=3$.)

Then
$$\mu(X) \leq \frac{1-\lambda}{1+\lambda + v \lambda^{u+1} + (q-2-v) \lambda^u}.$$

In particular, if $q\geq 5$
$$\mu(X) \leq \frac{1-\lambda}{q-4(1-\lambda)}.$$
\label{lem:onereg}
\end{lemma}
\begin{proof}

Without loss of generality, suppose $B_X(s_X)=1$,
$B'_X(s_X)=2$, and $c_1\geq c_2$.
Let $E=E(R_X)-s_X$, noting that $|E|\leq 3$. Let $n_i$ be the number
of edges in $E$ that are assigned spin~$i$ by $B_X$.
Note that
$c_i = \lambda^{n_i}$, so
the constraint $c_1\geq c_2$ just says
$n_2\geq n_1$.

Now we wish to choose~$B_X$ in order to maximise~$\mu(X)$, or,
equivalently, to minimise
$$Z = \frac{c}{(1-\lambda)c_1}.$$

First note that $n_1=0$ since $Z$ can be reduced by recolouring edges
coloured~$1$ with colour~$2$. Thus $c_1=1$.

Now we want to set $n_2,\ldots,n_q$ in order to
minimise~$c=\lambda^{n_3}+\cdots+\lambda^{n_q}$, where
$n_3+\cdots+n_q \leq 3$.
Since $\lambda \leq 1$, we want to take $n_3+\cdots + n_q = 3$.

Next, note that there is an optimal solution in which all $n_j$ and $n_k$ are
within $1$ of each other. To see this, consider a solution with $n_j>n_k+1$.
The boundary obtained by
reassigning one of the $j$~edges with spin~$k$ has a
$c$-value which is at least as small, since the new $c$-value minus the old one
is
$$-\lambda^{n_j} - \lambda^{n_k} + \lambda^{n_j-1} + \lambda^{n_k+1} =
(1-\lambda)(\lambda^{n_j-1}-\lambda^{n_k})\leq 0.$$
So the optimum value of $c$ is $v \lambda^{u+1} + (q-2-v) \lambda^u$, which
gives the first part of the lemma.

To derive the bound for $q\geq 5$ note
that for $q\geq 6$ we have $u=0$ and $v=3$.
For $q=5$ we have $u=1$ and $v=0$.
Both of these give the same bound.
\end{proof}

We now turn to the proof of Theorem~\ref{thm:no-computer}.
The cases ($q>7$), ($q=7,\lambda=0$) and ($q=6,\lambda=0$) follow from previous work
(see Proposition~\ref{prop:known}).
For each of the remaining cases we will use Lemma~\ref{lem:onereg}
to show that if $X$ is a size-$1$ boundary pair then $\mu(X)<1/3$.
This implies by Lemma~\ref{lemma:miss} that
every boundary pair $X$ satisfies $\nu(X)<1/3$ and
there is an $\epsilon>0$ so that
every boundary pair~$X$ satisfies
$$\nu(X)\leq{(1-\epsilon)}\frac13.$$ By induction on~$d$
(see Lemma~18 of~\cite{gmp}),
we get $\Gamma_d(X)\leq {(1-\epsilon)}^{d}$.
Hence, by Lemma~\ref{lem:tree} we have strong spatial mixing (and the theorem
is proved).

We now consider the remaining cases. The second part of Lemma~\ref{lem:onereg}
applies for $q\geq 5$ where
$q-4(1-\lambda)>3(1-\lambda)$, i.e., $\lambda>1-q/7$.
This finishes the cases with $q\geq 5$.

For $q=4$ we use the first part of Lemma~\ref{lem:onereg} with $u=1$ and $v=1$
and for $q=3$ we use the first part of Lemma~\ref{lem:onereg}
with $u=3$ and
$v=0$.

\begin{remark}
Lemma~\ref{lem:onereg} applies to the Potts model in a more general setting
than the one considered in this paper. In particular, it applies to the
Potts model on a general graph with maximum degree~$\Delta$.
In the generalised version, the ``$3$'' in the definition of $v$ and $u$ becomes
``$\Delta-1$''.
The final part of the lemma applies when $q\geq \Delta+1$.
It gives $\mu(X)\leq(1-\lambda)/(q-\Delta(1-\lambda))$,
so, for example,
we get the following result,
which is slightly better than the condition derived
by Salas and Sokal and discussed in Section~\ref{sec:context}.
 \end{remark}

\begin{theorem}
Consider the anti-ferromagnetic Potts model on a graph~$G$ with maximum degree~$\Delta$
with parameters~$q$ and $\lambda\leq 1$. There is strong spatial mixing if
$q>(1-\lambda)(2\Delta-1)$.
\end{theorem}

\section{Proof of Theorem~\ref{thm:computer} for $q=6$ and positive $\lambda$}\label{sec:comp}
\label{sec:six}

We will prove strong spatial mixing for $q=6$ and
$\lambda>0$\,\footnote{
The same proof technique applies to the $\lambda=0$ case. However we exclude $\lambda=0$
because the result is already known~\cite{six} and excluding $\lambda=0$ simplifies
our presentation.}
by showing that
there is an $\epsilon>0$ such that, for every boundary pair~$X$
and every $d\geq 1$, $\Gamma_d(X)\leq{(1-\epsilon)}^{d}$.
Then we apply Lemma~\ref{lem:tree}.
Following Goldberg, Martin and Paterson~\cite{gmp}, we will consider the
geometry of the lattice to derive a system of recurrences whose solution gives
the desired bound.

We start by considering some particular boundary pairs. In particular,
we will be interested in a boundary pair~$X$ such that $R_X$ is one of
the
seven regions
$Q_1$, $Q_2$, $Q_3$, $Q_4$, $Q_5$, $Q_6$, and $Q_7$ depicted below.
As before, we denote vertices as squares in the diagrams and $s_X$ is the edge
between $f_X$ and its lower neighbour. This edge is marked with a short line segment.
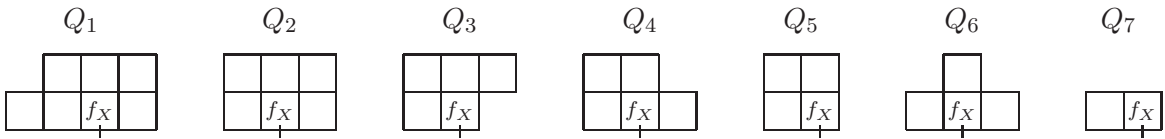
\begin{center}
\begin{tabular}{c p{0.5mm} c p{0.5mm} c p{0.5mm} c p{0.5mm} c p{0.5mm} c p{0.5mm} c}
$Q_1$ && $Q_2$ && $Q_3$ && $Q_4$ && $Q_5$ && $Q_6$ && $Q_7$ \medskip \\
\Qone && \Qtwo && \Qthree && \Qfour && \Qfive && \Qsix && \Qseven \\
\end{tabular}
\end{center}

\begin{lemma}
\label{lem:q_values6colours}
Suppose $q=6$ and $\lambda\in(0,1]$.
Let $p_1=41/118$, $p_2=179/501$, $p_3=79/216$,
$p_4=75/202$, $p_5=49/129$, $p_6=27/71$ and $p_7=3/7$. Define
$q_i = p_i + \delta$ for $i \in \{1, \dots, 7\}$ where $\delta = 1/1000$.
Suppose $X$ is a boundary pair with region $R_X = Q_i$ above.
Then $\mu(X) \leq q_i$.
\end{lemma}
\begin{proof}
The lemma is proved by computation.
For each region $Q_i$ we have considered every boundary
pair~$X$ which has $R_X = Q_i$.
Each such boundary pair consists of a pair $(B_X,B'_X)$ of boundary
configurations which differ
only on the edge~$s_X$, obeying the
requirements in Definition~\ref{def:bp}.
For each such boundary pair, we calculated
a rational function in~$\lambda$, $\mu_X(\lambda)$,
which gives an upper bound on $\mu(X)$ for any particular value of~$\lambda$.
The polynomials in the numerator and denominator of $\mu_X(\lambda)$ have
integer coefficients.
In order to find an upper bound on $\mu_X(\lambda)$ for $\lambda\in(0,1]$,
we partitioned the interval $[0,1]$ into smaller intervals $[a,b]$. We then
computed an upper bound for $\mu_X(\lambda)$
for $\lambda\in[a,b]$
by taking $\lambda = a$ for
negative terms in the numerator and $\lambda = b$ for positive terms in the
numerator. All terms in the denominator are positive so we use $\lambda = a$.
This computation was carried out exactly with no approximations.
Working through all  boundary pairs $X$  and an appropriate collection of intervals
$[a,b]$ we established the upper bounds given in the lemma.
\end{proof}

\begin{remark}
The value~$p_i$ defined in the statement of Lemma~\ref{lem:q_values6colours}
is defined by $$p_i =
\max_{X : R_X = Q_i} \mu_X(0).$$
$\mu_X(\lambda)$ is not monotonic in $\lambda$ in general.
A simple non-monotonic example is
the boundary pair consisting of a size-$1$ region with boundary $1,2,2$
where $s_X$ is assigned spins $1$ and $2$.
For this boundary pair, $c_1=\lambda$, $c_2=\lambda^2$ and $c_3=c_4=c_5=c_6=1$
so
$$\mu_X(\lambda)=\frac{(1-\lambda)c_1}{(1+\lambda)c_1+c}
=\frac{(1-\lambda)\lambda}{(1+\lambda)\lambda+4}.$$
Nevertheless,
$\max_{X: R_X = Q_i} \mu_X(\lambda)$ seems to be monotonically
decreasing in~$\lambda$.
\end{remark}

We now define some sets $V$, $W$, $U$, $T$, $S$, $R$ of boundary
pairs $X$. The sets depend only on the region $R_X$ and the edge $s_X$, but
not on the boundary configurations~$B_X$ and~$B'_X$. The following diagram
illustrates the sets.
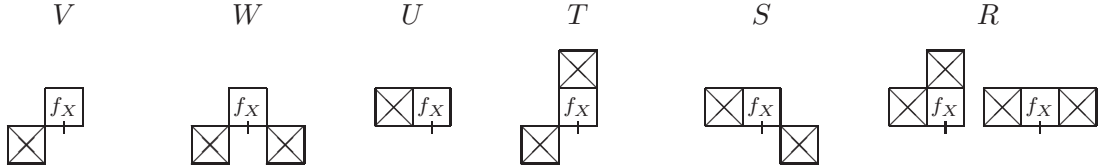
\begin{center}
\begin{tabular}{c p{1mm} c p{1mm} c p{1mm} c p{1mm} c p{1mm} c}
$V$ && $W$ && $U$ && $T$ && $S$ && $R$ \medskip \\
\Vfig && \Wfig && \Ufig && \Tfig && \Sfig && \Ronefig \Rtwofig
\end{tabular}
\end{center}
A crossed out square represents a vertex that is {\em not} in the region $R_X$.
Squares that are not drawn represent vertices that are either in, or not
in, the region $R_X$.
As before, the edge $s_X$ is marked with a short line segment.
The diagrams may be rotated according to the symmetries of $\mathbb{Z}^2$.
For example, a boundary pair $X$ belongs to the
set $R$ if at least two of the neighbours of $f_X$ are not in $R_X$.
A boundary pair $X$ belongs to the set $U$ if the left or right
neighbour of $f_X$ (or both) is not in $R_X$. Obviously these sets are
not disjoint.

We will now define some recurrences. Let $\Gamma_d$
denote the maximum, over boundary pairs~$X$, of $\Gamma_d(X)$.
Let $V_d$ denote the maximum of
$\Gamma_d(X)$ over boundary pairs $X\in V$ and we use similar
notation for the other sets.

Consider a  boundary pair~$X$.
We will consider six cases below. Every boundary pair is covered by
exactly one of the cases (up to symmetry).
In the diagrams, an empty square represents a vertex in the region~$R_X$.
As before, a crossed out square represents a vertex not in $R_X$,
and all other vertices can be either in~$R_X$ or not in~$R_X$.
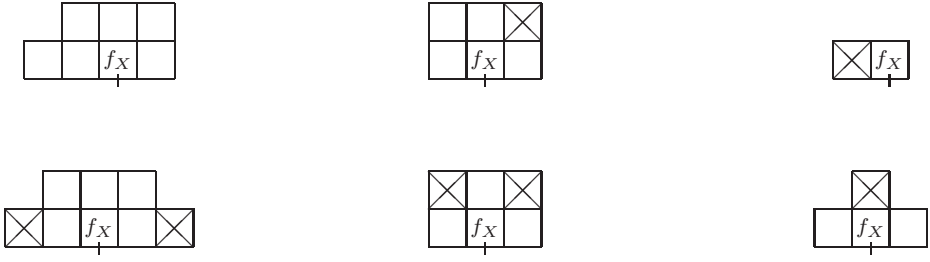
\begin{center}
\caseLowTop{\GammaOne}{\quad}
\caseLowTop{\GammaThree}{\quad}
\caseLowTop{\GammaFive}{\quad}
\caseLowNext{\GammaTwo}{\quad}
\caseLowNext{\GammaFour}{\quad}
\caseLowNext{\GammaSix}{\quad}
\end{center}

To see that the cases cover all boundary pairs, note the that the
left-most four diagrams cover all cases in which all three neighbours of $f_X$
are present. The lower central diagram applies if neither of the diagonal
vertices is present in~$R_X$. The diagram above that applies if just one of
the diagonal vertices is present. The two diagrams to the left apply if both of
the diagonal vertices are present.

We now add an inequality below each diagram giving an upper bound
on $\Gamma_d(X)$ for $d\geq 2$
when $X$ is a boundary pair covered by the corresponding case.
The inequality arises by considering the boundary pairs corresponding to the
children of~$X$ in the tree~$T_X$. The values $q_1$--$q_7$ are from
Lemma~\ref{lem:q_values6colours}.

\begin{center}
\caseLowTop{\GammaOne}{$\Gamma_d(X) \leq q_1(\Gamma_{d-1} + 2V_{d-1})$}
\caseLowTop{\GammaThree}{\mbox{$\Gamma_d(X) \leq q_4(V_{d-1} + U_{d-1} + S_{d-1})$}}
\caseLowTop{\GammaFive}{$\Gamma_d(X) \leq U_d$}
\caseLowNext{\GammaTwo}{$\Gamma_d(X) \leq q_2(\Gamma_{d-1} + 2T_{d-1})$}
\caseLowNext{\GammaFour}{$\Gamma_d(X) \leq q_6(2S_{d-1} + R_{d-1})$}
\caseLowNext{\GammaSix}{$\Gamma_d(X) \leq q_7(2W_{d-1})$}
\end{center}
For example, consider a boundary pair~$X$ covered by the
lower centre diagram.
We will now show how to prove $\Gamma_d(X) \leq q_6(2S_{d-1} + R_{d-1})$.
In the construction of~$T_X$, for every pair of spins $c\in Q$, $c'\in Q$,
$c\neq c'$ we introduce a child $r_{c,c'}$ of the root~$r$.
We construct three boundary pairs $X_1(c,c')$ (where the new distinguished
edge goes left from~$f_X$), $X_2(c,c')$ (where the new distinguished edge goes
up from~$f_X$) and $X_3(c,c')$ (where the new distinguished edge goes
right from~$f_X$). The boundary pair $X_1(c,c')$ is in~$S$
(this can be verified by consulting the diagram corresponding to~$S$ above),
so $\Gamma_{d-1}(X_1(c,c'))\leq S_{d-1}$. Similarly,
$X_3(c,c')\in S$, so $\Gamma_{d-1}(X_3(c,c'))\leq S_{d-1}$.
Finally, $X_2(c,c')\in R$ (this can be verified by consulting the
diagram corresponding to $R$ above), so $X_2(c,c')\leq R_{d-1}$.
Since $\nu(X)$ is the sum of the probabilities $p_X(c,c')$, we conclude
that $\Gamma_d(X)\leq \nu(X)(2 S_{d-1}+R_{d-1})$. Now we apply Lemma~\ref{lemma:miss}
and Lemma~\ref{lem:q_values6colours}
to get $\nu(X)\leq \mu(X) \leq q_6$.
Thus, we have shown $\Gamma_d(X) \leq q_6(2S_{d-1} + R_{d-1})$.
The other inequalities are derived similarly.

Putting all six cases together, we get the following inequality for
$d \geq 2$.
\begin{align}
\Gamma_d \leq \max(&q_1(\Gamma_{d-1} + 2 V_{d-1}), \nonumber \\
&q_2(\Gamma_{d-1} + 2 T_{d-1}), \nonumber \\
&q_4(V_{d-1} + U_{d-1} + S_{d-1}), \nonumber \\
&q_6(2 S_{d-1}+R_{d-1}), \nonumber \\
&U_d, \nonumber \\
&q_7(2 W_{d-1})). \label{eq:gamma}
\end{align}

By  re-considering similar scenarios with the additional assumption
that $X \in V$ we derive a corresponding upper bound for $V_d$. The
following cases cover all  boundary pairs in $V$.

\bigskip
\begin{center}
\caseTallTop{\VOneA \quad \VOneB}{
\mbox{$\Gamma_d(X) \leq q_1(U_{d-1} + \Gamma_{d-1} + V_{d-1})$}}
\caseTallTop{\VFiveA \quad \VFiveB}{$\Gamma_d(X) \leq U_d$}
\caseTallTop{\VTwo}{
\mbox{$\Gamma_d(X) \leq q_2(R_{d-1} + \Gamma_{d-1} + T_{d-1})$}}
\caseTallNext{\VThree}{$\Gamma_d(X) \leq q_4(2U_{d-1} + S_{d-1})$}
\caseTallNext{\VFour}{$\Gamma_d(X) \leq q_6(S_{d-1} + 2R_{d-1})$}
\caseTallNext{\VSix}{$\Gamma_d(X) \leq q_7(S_{d-1} + W_{d-1})$}
\caseTallNext{\VSeven}{\mbox{$\Gamma_d(X) \leq q_4(V_{d-1} + U_{d-1} + R_{d-1})$}}
\end{center}
\bigskip

Putting these together, we get this inequality for $d \geq 2$.
\begin{align}
V_d \leq \max(&q_1(U_{d-1} + \Gamma_{d-1} + V_{d-1}), \nonumber \\
&q_2(R_{d-1} + \Gamma_{d-1} + T_{d-1}), \nonumber \\
&q_4(2 U_{d-1} + S_{d-1}), \nonumber \\
&q_6(S_{d-1} + 2 R_{d-1}), \nonumber \\
&q_4(V_{d-1} + U_{d-1} + R_{d-1}), \nonumber \\
&U_d, \nonumber \\
&q_7(S_{d-1} + W_{d-1})). \label{eq:V}
\end{align}

In a similar manner we can find an upper bound for $W_d$, and the following
cases cover the  boundary pairs in $W$.
\bigskip
\begin{center}
\caseTallTop{\WOne}{$\Gamma_d(X) \leq q_6(2U_{d-1} + \Gamma_{d-1})$}
\caseTallTop{\WThree}{$\Gamma_d(X) \leq q_7(2S_{d-1})$}
\caseTallTop{\WTwo}{$\Gamma_d(X) \leq U_d$}
\end{center}
\bigskip
These cases give the following inequality for $d \geq 2$.
\begin{equation}
\label{eq:W}
W_d \leq \max(q_6(2 U_{d-1}+ \Gamma_{d-1}), U_d, q_7 (2 S_{d-1})).
\end{equation}

The following
cases cover all  boundary pairs in $U$, so we can find an
upper bound for $U_d$.
\bigskip
\begin{center}
\caseLowTop{\UOne}{$\Gamma_d(X) \leq q_3(2V_{d-1})$}
\caseLowTop{\UTwo}{$\Gamma_d(X)  \leq q_5(V_{d-1} + U_{d-1})$}
\caseLowTop{\UThree}{$\Gamma_d(X) \leq q_7(S_{d-1} + U_{d-1})$}
\caseLowNext{\UFourA \quad \UFourB}{$\Gamma_d(X) \leq R_d$}
\end{center}
\bigskip
These give an upper bound on $U_d$ for $d \geq 2$.
\begin{equation}
\label{eq:U}
U_d \leq \max(q_3(2 V_{d-1}), q_5(V_{d-1} + U_{d-1}), q_7(S_{d-1} + U_{d-1}), R_d).
\end{equation}

The following cases illustrate the situation for
boundary pairs in $S$.
\bigskip
\begin{center}
\caseTallTop{\SOneA \quad \SOneB}{$\Gamma_d(X) \leq R_d$}
\caseTallTop{\STwo}{$\Gamma_d(X) \leq q_5(U_{d-1} + V_{d-1})$}
\caseTallTop{\SThree}{$\Gamma_d(X) \leq q_7(R_{d-1} + S_{d-1})$}
\end{center}
\bigskip
These give the following inequality for $d \geq 2$.
\begin{equation}
\label{eq:S}
S_d \leq \max(R_d, q_5 (U_{d-1}+V_{d-1}), q_7 (R_{d-1}+S_{d-1})).
\end{equation}

Now we derive a corresponding upper bound for $T_d$. The following
cases cover all  boundary pairs in $T$ (apart from those in $R$).
\bigskip
\begin{center}
\caseTallTop{\TOne}{$\Gamma_d(X) \leq q_7(W_{d-1} + S_{d-1})$}
\caseTallTop{\TTwo}{$\Gamma_d(X) \leq q_7(W_{d-1})$}
\caseTallTop{\TThree}{$\Gamma_d(X) \leq q_7(S_{d-1})$}
\end{center}
\bigskip
These give the following inequality for $d \geq 2$.
\begin{equation}
\label{eq:T}
T_d \leq \max(R_d,q_7(W_{d-1} + S_{d-1})).
\end{equation}

Finally, we derive an upper bound for $R_d$. The
following cases cover all  boundary pairs in $R$.
Notice that the middle diagram below does not exactly match the
set $Q_7$, but clearly we can use the value of $q_7$ to bound $\mu(X)$
also for this case.
\bigskip
\begin{center}
\caseLowTop{\ROneA \quad \ROneB}{$\Gamma_d(X) \leq q_7(W_{d-1})$}
\caseLowTop{\RTwo}{$\Gamma_d(X)=0$ for $d \geq 2$}
\end{center}
\bigskip
These give the following inequality for $d \geq 2$.
\begin{equation}
\label{eq:R}
R_d \leq \max(0,q_7 W_{d-1}).
\end{equation}

We now set $\epsilon=1/1000$ and show that for every $d\geq 1$, $\Gamma_d
\leq {(1-\epsilon)}^d$.
We define some rational numbers. Let $u = s = t = r = 7/10$
and $v = w = 92/100$.
We will prove by induction on $d$ that $\Gamma_d \leq {(1-\epsilon)}^d$,
$V_d\leq v {(1-\epsilon)}^d$, $W_d \leq w
{(1-\epsilon)}^d$, $U_d \leq u{(1-\epsilon)}^d$,
$S_d \leq s{(1-\epsilon)}^d$, $T_d \leq t {(1-\epsilon)}^d$,
and $R_d \leq r {(1-\epsilon)}^d$.

The base case is $d=1$. For any  boundary pair~$X$
we have $\Gamma_1(X) \leq \nu(X) \leq \mu(X)$ and from
Lemma~\ref{lem:onereg}
$$\mu(X) \leq\frac{1-\lambda}{6-4(1-\lambda)} \leq \frac12.$$
The base case then follows from the fact that
$$\frac{1}{2} \leq \min(1, v, w, u, s, t, r) (1-\epsilon).$$

The inductive step follows from the Equations~(\ref{eq:gamma}),
(\ref{eq:V}), (\ref{eq:W}), (\ref{eq:U}), (\ref{eq:S}), (\ref{eq:T})
and~(\ref{eq:R}).

First, we use Inequality~(\ref{eq:R}),
the facts that $r\geq0$ and $\epsilon\leq 1$ (so $0\leq r{(1-\epsilon)}^d$),
and the fact that $q_7 w \leq r(1-\epsilon)$ to show
$R_d \leq r{(1-\epsilon)}^d$.
Similarly, we use the inductive hypothesis, Inequality~(\ref{eq:T})
and the facts that $r\leq t$ and $q_7(w+s)\leq t(1-\epsilon)$ to show
$T_d \leq t{(1-\epsilon)}^d$.

Next, we establish upper bounds on $S_d$ and $U_d$. To show
$S_d\leq s {(1-\epsilon)}^d$, we use the inductive hypothesis
and Inequality~(\ref{eq:S}) together with the upper bound
$R_d \leq r{(1-\epsilon)}^d$ and the following facts:
$r\leq s$, $q_5(u+v)\leq s(1-\epsilon)$, and $q_7(r+s)\leq s(1-\epsilon)$.
To show $U_d\leq u {(1-\epsilon)}^d$, we use the inductive hypothesis
and Inequality~(\ref{eq:U}) together with the upper bound
$R_d \leq r{(1-\epsilon)}^d $ and the following facts:
$r\leq u$, $q_3 2 v\leq (1-\epsilon)u$, $q_5(v+u) \leq (1-\epsilon)u$,
and $q_7(s+u)\leq (1-\epsilon)u$.

Finally, we establish upper bounds on $W_d$, $V_d$ and $\Gamma_d$. All
of these bounds use the inductive hypothesis and the upper bound
$U_d \leq u{(1-\epsilon)}^d$ along with $u \leq w$, $u \leq v$ and $u \leq 1$.
To establish $W_d \leq w {(1-\epsilon)}^d$, we use Inequality~(\ref{eq:W})
along with the following facts: $q_6(2u+1) \leq (1-\epsilon)w$ and
$q_7 2 s \leq (1-\epsilon)w$. To establish $V_d \leq v {(1-\epsilon)}^d$,
we use Inequality~(\ref{eq:V}) along with the following facts:
\begin{align*}
q_1(u+1+v) &\leq v(1-\epsilon), \\
q_2(r+1+t) &\leq v(1-\epsilon), \\
q_4(2u+s)  &\leq v(1-\epsilon), \\
q_6(s+2r)  &\leq v(1-\epsilon), \\
q_4(v+u+r) &\leq v(1-\epsilon), \\
q_7(s+w)   &\leq v(1-\epsilon).
\end{align*}

Finally, to establish $\Gamma_d \leq {(1-\epsilon)}^d$, we use
Inequality~(\ref{eq:gamma}) along with the following facts:
\begin{align*}
q_1(1+2 v) &\leq 1-\epsilon, \\
q_2(1+2 t) &\leq 1-\epsilon, \\
q_4(v+u+s) &\leq 1-\epsilon, \\
q_6(2s+r)  &\leq 1-\epsilon, \\
q_7(2 w)   &\leq 1-\epsilon.
\end{align*}

This concludes the proof of Theorem~\ref{thm:computer} for $q=6$.

\section{Proof of Theorem~\ref{thm:computer} for $q=5$, $q=4$ and $q=3$}
\label{sec:other_colours}

The proof is the same as the proof for $q=6$ in Section~\ref{sec:six} except that for each
value of~$q$, we compute new values for
$q_1, \dots, q_7$ (as in Lemma~\ref{lem:q_values6colours}).
To find sufficiently small values we
need to constrain the value of $\lambda$. If $\lambda$ is too small
the values of $q_1, \dots, q_7$ get too large.
We do not repeat the values of~$\lambda$ already covered by Theorem~\ref{thm:no-computer}.

\begin{lemma}
\label{lem:q_values5colours}
Suppose $q=5$ and $\lambda \in [0.127, 0.286]$.
Let $p_1=7/20$, $p_2=9/25$, $p_3=19/50$,
$p_4=2/5$, $p_5=2/5$, $p_6=2/5$ and $p_7=1/2$.  Define
$q_i = p_i + \delta$ for $i \in \{1, \dots, 7\}$ where $\delta = 1/1000$.
Suppose $X$ is a  boundary pair with region $R_X = Q_i$.
Then $\mu(X) \leq q_i$.
\end{lemma}
\begin{lemma}
\label{lem:q_values4colours}
Suppose $q=4$ and $\lambda \in [0.262, 0.373]$.
Let $p_1=7/20$, $p_2=19/50$, $p_3=19/50$, $p_4=19/50$,
$p_5=2/5$, $p_6=19/50$ and $p_7=1/2$.  Define
$q_i = p_i + \delta$ for $i \in \{1, \dots, 7\}$ where $\delta = 1/1000$.
Suppose $X$ is a  boundary pair with region $R_X = Q_i$.
Then $\mu(X) \leq q_i$.
\end{lemma}
\begin{lemma}
\label{lem:q_values3colours}
Suppose $q=3$ and $\lambda \in  [0.393, 0.474]$.
Let $p_1=873/2500$, $p_2=9/25$, $p_3=48/125$,
$p_4=9/25$, $p_5=39/100$, $p_6=37/100$ and $p_7=1/2$.  Define
$q_i = p_i + \delta$ for $i \in \{1, \dots, 7\}$ where $\delta = 1/1000$.
Suppose $X$ is a  boundary pair with region $R_X = Q_i$.
Then $\mu(X) \leq q_i$.
\end{lemma}
\begin{remark}
Unlike Lemma~\ref{lem:q_values6colours}, the values of $p_i$ in the
lemmas above are strict upper bounds on $\max_{X: R_X = Q_i} \mu_X({\lambda})$,
where $\lambda$ is the smallest value in the specified intervals above.
Writing the exact values of $\max_{X: R_X = Q_i} \mu_X(\lambda)$ would require
many more digits. Again, these values seem to be monotonically decreasing in~$\lambda$.
\end{remark}
We use computation in the same manner as for the proof of Lemma~\ref{lem:q_values6colours}
to prove these lemmas. Following the proof of
the $q=6$ case of the theorem
and using the values of $q_i$ in the lemmas above, we can then
define new rational numbers $v$, $w$, $u$, $t$, $s$ and $r$, and prove
$\Gamma_d(X) \leq {(1-\epsilon)}^d$.

\end{document}